\documentclass[10pt,twocolumn,twoside]{IEEEtran}
\usepackage[utf8]{inputenc}
\usepackage[english]{babel}
                     
\usepackage{array}
\newcolumntype{P}[1]{>{\centering\arraybackslash}p{#1}}
\usepackage{float}
\usepackage{tabu}
\usepackage{amsmath}
\usepackage{mathtools}

\usepackage{blindtext}
\usepackage{multicol}

\usepackage{graphics} 
\usepackage{mathptmx} 
\usepackage{times} 
\usepackage{amssymb}  
\usepackage{multirow}
\usepackage{subcaption}
\usepackage{csquotes}
\usepackage{algpseudocode}
\usepackage{algorithm}
\usepackage{enumerate}

\usepackage{nicefrac}
\usepackage{revsymb}
\usepackage{cite}
\usepackage{tikz}
\usetikzlibrary{shapes.geometric, arrows}
\usepackage{longtable}
\usepackage{amsbsy}
\usepackage{authblk}
\usepackage{leftidx}
\usepackage{ragged2e}
\usepackage[font={scriptsize}]{caption}
\usepackage{amsthm}
\newtheorem{theorem}{Theorem}
\newtheorem{lemma}{Lemma}
\newtheorem{remark}{Remark}
\newtheorem{proposition}{Proposition}

\let\originaleqref\eqref
\renewcommand{\eqref}{\originaleqref}

\makeatletter
\newcommand*\titleheader[1]{\gdef\@titleheader{#1}}
\AtBeginDocument{%
  \let\st@red@title\@title
  \def\@title{%
    \bgroup\normalfont\large\centering\@titleheader\par\egroup
    \vskip1.5em\st@red@title}
}
\makeatother

\titleheader{This work has been accepted for publication in IEEE Transactions on Control of Network Systems.}

\title{Adversary Detection and Resilient Control for Multi-Agent Systems}

\begin{document}

\author {Aquib Mustafa and Dimitra Panagou \\
\thanks{This work was partially sponsored by the Office of Naval Research (ONR), under grant number N00014-20-1-2395. The views and conclusions contained herein are those of the authors only and should not be interpreted as representing those of ONR, the U.S. Navy or the U.S. Government. This work was also partially supported by the National Science Foundation (NSF) under award number 1931982.}

\thanks{Aquib Mustafa and Dimitra Panagou are with the Department
of Aerospace Engineering, University of Michigan, Ann Arbor, MI-48109 
USA (e-mails: aquibm@umich.edu, dpanagou@umich.edu).}
}

\maketitle
\thispagestyle{empty}
\pagestyle{empty}
 \begin{abstract}
This paper presents an adversary detection mechanism and a resilient control framework for multi-agent systems under spatiotemporal constraints. Safety in multi-agent systems is typically addressed under the assumption that all agents collaborate to ensure the forward invariance of a desired safe set. This work analyzes agent behaviors based on certain behavior metrics, and designs a proactive adversary detection mechanism based on the notion of the critical region for the system operation. In particular, the presented detection mechanism not only identifies adversarial agents, but also ensures all-time safety for intact agents. Then, based on the analysis and detection results, a resilient QP-based controller is presented to ensure safety and liveness constraints for intact agents. Simulation results validate the efficacy of the presented theoretical contributions.
\end{abstract}


\IEEEpeerreviewmaketitle

\vspace{-0.2cm}
\section{Introduction}

In recent years, research for safety-critical systems has received vast recognition as safety is one of the prime requirements for autonomous systems. For a given system, safety is accomplished by ensuring forward invariance of a safe set, which is a subset of the system's state space. The objective is to design a control law such that the closed-loop system trajectories remain always in the safe set. In the existing literature, control barrier function (CBF) based approaches that leverage quadratic programming (QP) \cite{ames2019control, ames2016control,srinivasan2018control,garg2019control,lindemann2020barrier} methods have shown impactful results for providing safety guarantees for both single-agent \cite{ames2019control,ames2016control,xiao2019control} and multi-agent systems \cite{wang2017safety,glotfelter2017nonsmooth,guerrero2019realization,lindemann2020barrier}. These approaches are well-suited for online implementation as QPs can be efficiently solved in real-time \cite{glotfelter2017nonsmooth,guerrero2019realization, wang2017safe, pickem2017robotarium}.


The aforementioned results generally consider that all agents behave normally, i.e., they apply the nominally-specified control actions. However, these systems are vulnerable to a variety of adversaries, which aim to intentionally violate desired safety or goal-reaching constraints for normally-behaving agents within given control constraints. Therefore, it is of vital importance to design a proactive adversary detection mechanism and resilient control framework that can mitigate the effect of adversarial agents while ensuring all-time safety for intact agents.

In the existing literature, several remarkable results for resilient control are presented for multi-agent systems \cite{pasqualetti2011consensus,Sundaram2011,usevitch2019resilient,saulnier2017resilient,zhou2018resilient,guerrero2019realization,mustafa2019attack,pirani2019design}. In particular, mean-subsequence-reduced (MSR) based resilient control protocols for multi-agent systems are presented in\cite{pasqualetti2011consensus,Sundaram2011,usevitch2019resilient}. Resilient algorithms for flocking and active target tracking applications are presented in  \cite{saulnier2017resilient,zhou2018resilient}, respectively. An adaptive control and game-theoretic resilient designs are presented in \cite{mustafa2019attack} and \cite{pirani2019design} to directly mitigate the effect of adversaries without identifying them.  However, to our knowledge, no prior studies using CBF approaches have considered identification of adversarial agent and resilient design for multi-agent CBF with safety and goal reaching objectives. Recently, authors in \cite{usevitch2021adversarial} presented adversarial resilience for sampled-data systems under safety constraints without any adversary identification. In \cite{guerrero2019realization}, authors used CBFs to ensure that the communication topology satisfies the r-robustness property in finite time. However, the robots in formation are assumed to apply the nominal CBF-based controller without any adversarial misbehavior. {Similarly, in \cite{clark2020control} a class of fault-tolerant stochastic CBF is presented that provide probabilistic guarantees on the safety.  This work is mainly focused on solving a secure state estimation problem by deriving geometrical conditions to resolve conflicts between the constraints that may arise due to sensor faults or attacks.}

{In this paper, we present an adversary detection mechanism as well as a resilient CBF framework for multi-agent systems. We consider heterogeneous multi-agent systems, modeled by control-affine dynamics, where in the presence of adversarial agents, intact agents are subject to accomplish the following objectives: (i) remain inside a safe set, which can be in general time-varying, (ii) reach desired goal locations either individually or in formation, and (iii) proactively identify adversarial agents in their neighborhood and take resilient action to ensure all-time safety. To achieve these objectives, this work first analyzes agent behaviors based on metrics that act as real-time behavior monitors, and then designs a proactive adversary detection mechanism based on the notion of the critical time and critical zone for the system operation. In particular, the presented critical zone is evaluated over the critical time window under best and worst-case control actions corresponding to intact and adversarial agents, respectively. Then, the augmentation of the critical zone with the desired safety constraints provides robustness to the agent's safe set such that the presented detection mechanism not only identifies adversarial agents but also acts to ensure all-time safety for intact agents. Finally, based on the presented behavior analysis and proactive adversary detection, a resilient QP-based controller is designed to ensure all-time safety for intact agents, in the presence of adversarial agents. The overall architecture is shown in Figure 1.}

The rest of this paper is organized as follows. First, Section II provides the notations and Section III presents the problem formulation. Then, Section IV formulates the behavior analysis and detection mechanism. Next, the resilient CBF mechanism is presented in Section V and simulation results are provided in Section VI. Finally, concluding remarks are discussed in Section VII.

\vspace{-0.32cm}

\section{Notations}

$\mathbb{R}$ and  $\mathbb{R_{+}}$ represent the sets of real numbers and non-negative real numbers, respectively. $\mathbb{R}^n$ denotes $n$-dimensional Euclidean space. $\left\| x \right\|$ denotes Euclidean norm of vector $x \in {{\mathbb{R}}^{n}}$. The set of integers greater than $m$ is represented by $ \mathbb{Z}_{>m}$. The superscript ${(.)^T}$ denotes transposition. The cardinality of a set $S$ is denoted by $|S|$. The Lie derivative of a continuously differentiable function $V: \mathbb{R}^{n} \to \mathbb{R}$ along a vector $f: \mathbb{R}^{n} \to \mathbb{R}^{n}$ at point $x \in {{\mathbb{R}}^{n}}$ is represented as $L_{f}V(x) \triangleq  \frac{\partial V(x)}{\partial x}f(x).$ We use $\partial S$ to denote the boundary of a closed set $S$ and $int(S)$ to denote its interior. ${diag}\left( {{A_1},\ldots,{A_n}} \right)$ represents a diagonal matrix with ${A_i}$ as its diagonal entries, $\forall, i \in [1, \dots,n]$. $\wedge$ or $\bigcap$ denotes conjunction/and operator. Eventual and global temporal operators are represented by $\diamondsuit$ and $\square$.

\vspace{-0.3cm}

\section{Problem Formulation}
Consider a group of $ N \in \mathbb{Z}_{>0}$ agents, with the set of agents represented by $\mathcal{V}$ and each agent indexed $\{1,\dots,N\}$. The system dynamics of each agent $i \in \mathcal{V}$ is given by
\vspace{-0.1cm}
\begin{equation} \label{eq1} 
\dot{x}_{i} (t)=f_{i} (x_{i} (t))+g_{i} (x_{i} (t))u_{i} (t),
\vspace{-0.1cm}
\end{equation} 
 where the state vector is $x_{i}(t)=[p_i(t) \,\,\,\,\,\varphi_i(t)] \in \mathbb{R}^{3}, $ with  $p_i(t) \in \mathbb{R}^{2}$ and $\varphi_i(t) \in \mathbb{R}$ denoting the position vector and orientation, respectively, of agent $i$ with respect to a global reference frame. The vector $u_{i}(t) \in \mathbb{R}^{m_i} $ denotes the control input of agent $i$. The functions $f_{i} \in \mathbb{R}^{3} $ and $g_{i} \in \mathbb{R}^{{3} \times {m_i}}$ may differ among agents, but are all locally Lipschitz. {We denote $f_i^p \in \mathbb{R}^{2}$ and $g_i^p \in \mathbb{R}^{{2} \times {m_i}}$ the sub-matrices of $f_i$ and $g_i$ corresponding to the position-vector dynamics in \eqref{eq1}.} The control input constraints for each input $u_{i}(t)$ are represented by a nonempty, convex, compact polytope, i.e., $u_{i}(t) \in {\rm {\mathcal U}}_{i} (x_{i}(t) )=\{ u\in \mathbb{R}^{m_{i} } :A_{i}(x_{i}(t) )u\le b_{i} (x_{i}(t) )\} $ where the functions $A_{i}(x_{i}(t)) :\mathbb{R}^{3 } \to \mathbb{R}^{q_{i} \times m_{i} } $ and $b_{i}(x_{i}(t))  :\mathbb{R}^{3} \to \mathbb{R}^{q_{i} }$ are locally Lipschitz on their respective domains. Moreover, the collection of the position vectors and the control input vectors are represented as  $\vec{p}=[p_{1}^{T}, \,p_{2}^{T},\,\dots,\,p_{N}^{T}]^T \in \mathbb{R}^{{2N}}$ and $\vec{u}=[u_{1}^{T}, \,u_{2}^{T},\,\dots,\,u_{N}^{T}]^T \in \mathbb{R}^{\underbar{m}}$ with $\underbar{m}= \sum\nolimits_{i = 1}^N {{m_i}} $,  respectively. 
 
 For each agent $i$, the conjunction of $m$ different safety constraints $h_n^i:\mathbb R^{2N}\rightarrow \mathbb R$, $n\in\{1,\dots,m\}$, is represented by the composite control barrier function (CBF) $h_{i}^{s}(\vec{p}):\mathbb{R}^{2N} \to \mathbb{R}^{}$ via Boolean AND operations using the log-sum-exp (LSE) smooth approximation to the max(.) function \cite{lindemann2020barrier}, given by

\vspace{-0.0cm}
\noindent
 \begin{equation}\label{eq0}
h_{i}^{s}(\vec{p})= LSE[h_1^i,h_2^i,\dots,h_m^i]=ln(\sum_{n=1}^{m} exp^{h_n^i}).
 \end{equation}
  
 \noindent

 \begin{figure}[!t]
\centering{\includegraphics[width=0.95\columnwidth] {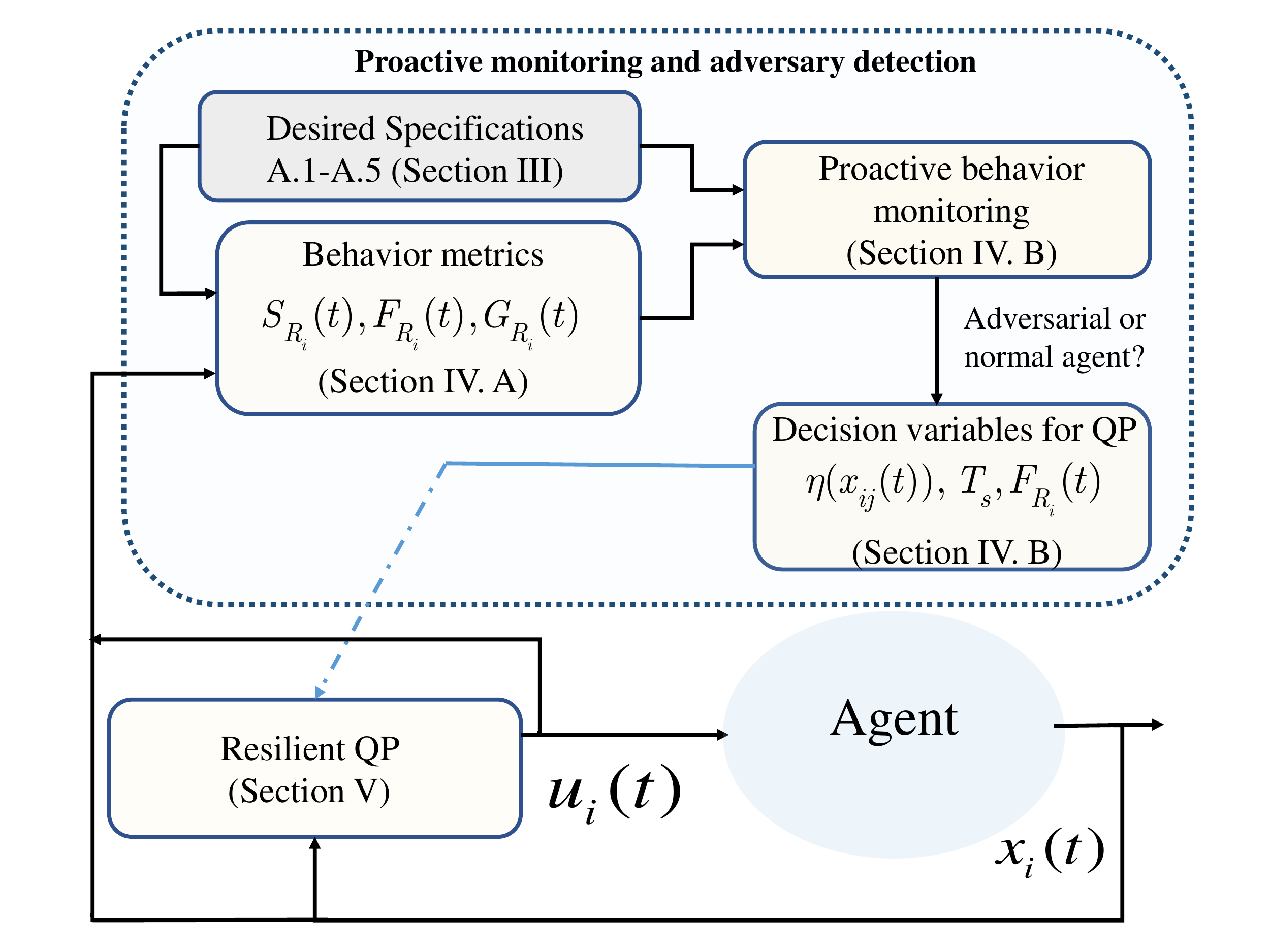}}
 \caption{Agents behavior: for desired specification without any adversary.}
\label{fig2B}
\vspace{-0.4cm}
\end{figure}

 
{We consider a set $S_{i}^{s}$ defined as the sublevel set of a continuously differentiable function $h_{i}^{s}(\vec{p})$, given by $S_i^s = \{ \vec p|h_i^s(\vec p) \le 0\}, \,\,\,
\partial S_i^s = \{ \vec p|h_i^s(\vec p) = 0\}, \,\,\,
{\mathop{\rm int}} (S_i^s) = \{ \vec p|h_i^s(\vec p) < 0\}.$ In this paper, we refer $S_i^s$ as a safe set for agent $i$, and assume that forward invariance of this set $S_i^s$ can always be ensured \cite{ames2019control}, see Lemma 1 later on.} We consider two class of safety constraints and their corresponding safe sets, namely, inter-agent and agent-to-obstacle safety constraints. We represent the conjunction of inter-agent safety constraints  as $ \bar h_i^s(\vec p) = \bigcap\nolimits_{j \in \mathcal{N}_i} { h_{ij}^{s}} ({p_i},\,{p_j})$ with 
\vspace{-0.12cm}
  \begin{equation}\label{s1}
 h_{ij}^{s} (p_{i},\,p_{j} ) =d -\left\| p_{i} -p_{j} \right\| \le 0,
 \vspace{-0.12cm}
 \end{equation}
 where $d$ is a desired inter-agent safety distance, and the set ${\rm {\mathcal N}}_{i} =\{ j|\, \left\| p_{i} -p_{j} \right\| \le R_{s} \} $ is the set of neighbors of agent $i$, where the limited sensing radius $R_{s} >d$. Then, the inter-agent safe set is defined as 
 \vspace{-0.25cm}
\begin{equation}\label{s2}
\bar S_{i}^{s} =\{ \vec{p}|\bar h_{i}^{s}(\vec{p}) \le 0\},
\vspace{-0.1cm}
 \end{equation}
with  $\bar S_{i}^{s} \supset S_{i}^{s}$. Similarly, the conjunction of agent-to-obstacle safety constraints is denoted by $\hat{h}_i^s({p_i}) = \bigcap\nolimits_{{o_j} \in {{\cal O}_i}} {h_i^{{o_j}}} ({p_i})$ with
\vspace{-0.15cm}
\begin{equation}\label{s3}
h_i^{o_j} (p_{i})={r_{{o_j}}} - \left\| {{p_i} - {c_{{o_j}}}} \right\| \leq 0,
\vspace{-0.15cm}
 \end{equation}
{where $c_{o_j}$ and $r_{o_j}$ denote the center and radius of spherical obstacles, and ${\mathcal{O}_i}$ denotes the set of the obstacles for the agent $i$.} Then, the agent-to-obstacle safe set is defined as
\vspace{-0.15cm}
 \begin{equation}\label{s4}
 \hat{S}_{i}^{s} =\{{p_i}|\hat{h}_i^s({p_i}) \le 0\},
\vspace{-0.15cm}
 \end{equation}

 \noindent
 with  $\hat S_{i}^{s} \supset S_{i}^{s}$.
  Moreover, all pairwise safety constraints, i.e., inter-agent and agent-to-obstacle safety constraints can be encoded in the augmented CBF $h_{i}^{s}(\vec{p})$ in \eqref{eq0} with
  \vspace{-0.15cm}
\begin{equation}\label{s5}
S_i^s = \bar S_i^s\cap \hat {S}_i^s,
\vspace{-0.15cm}
\end{equation}
where $m = \left| {{{\mathcal N}_i}} \right| + \left| {{{\mathcal O}_i}} \right|$ denotes the total number of safety constraints. {Note that while in principle one can define a CBF $h_i^s$ that is a function of both position and orientation, here we chose to define the inter-agent and agent-to-obstacle safety constraints in terms of the Euclidean distance only, as they encode that the area footprints of the agents should never intersect for any possible orientations. Note also that, for the dynamics in (1), the considered control action constraint in the QP-based control design directly provides a constraint on the rate of change of orientation.}


Next, we present some preliminaries related to the performance of the agents. 

\smallskip

\noindent
\textbf{Definition 1.} A continuously differentiable function $V_{i}^{g}(\vec{p}) :\mathbb{R}^{2N} \to \mathbb{R}^{}$ is called exponentially stabilizing control Lyapunov function (ES-CLF) for {the positional dynamics of} \eqref{eq1} if there exist $\beta_1, \beta_2, \beta_3 \in \mathbb{R_{+}} $ such that following conditions hold
\begin{equation}
\begin{array}{c}
{\beta _1}{\left\| {\vec p} \right\|^2} \le V_i^g(\vec p) \le {\beta _2}{\left\| {\vec p} \right\|^2}\\
\mathop {\inf }\limits_{{u_i}(t) \in {\cal U}_{i}} [{{L}_{{f_i^p}}}V_i^g(\vec p) + {{L}_{{g_i^p}}}V_i^g(\vec p){u_i}(t) + {\beta _3}V_i^g(\vec p)] \le 0.{\kern 1pt} 
\end{array} 
\end{equation}
{where $f_i^p \in \mathbb{R}^{2}$ and $g_i^p \in \mathbb{R}^{{2} \times {m_i}}$ denotes the sub-matrix of $f_i$ and $g_i$ corresponding to position vector dynamics in \eqref{eq1}.}


Reaching a goal location for agent $i$ can be encoded via the following candidate Lyapunov function
\vspace{-0.2cm}
\begin{equation}\label{s6}
V_{i}^{g}(p_i)= \left\| p_i - G_i \right\|^2,\,\, \forall i \in \mathcal{V},
\end{equation}
where $G_i \in \mathbb{R}^{2}$ denotes a goal point or goal region. Similarly, the goal reaching for a collaborative task among the set of agents ${{\cal V}_f} \subseteq{\cal V}$ in the form of formation can be encoded as 
\vspace{-0.2cm}
\begin{equation}\label{s6a}
    \bar{V}_{f}^{g}(\bar{p})= \left\| \bar{p} - G_f \right\|^2,
    \vspace{-0.1cm}
\end{equation}
where $G_f \in \mathbb{R}^{2}$ denotes a goal point or goal region and $\bar{p} = \frac{1}{{\left| {{{\cal V}_f}} \right|}}\sum\limits_{i \in {{\cal V}_f}} {{p_i}}$. We also consider that agents need to maintain a formation over time such that 
\vspace{-0.15cm}
\begin{equation}\label{c4}
 h_i^{{{\cal F}_i}}(\vec p (t))=  \mathop {\lim }\limits_{t \to \infty } \left\| {{p_i}(t) - p_i^*(t)} \right\| \to 0
,\,\,\forall i \in {{\cal V}_f},
\end{equation}
where  $p_i^*(t) = \frac{1}{{\left| {{{\cal N}_i}} \right|}}\sum\limits_{j \in {{\cal N}_i}} {({p_j}(t) + {c_{ji}})}$ and $c_{ij}$ denotes an inter-agent distance for the formation.

\smallskip

\noindent
\textbf{Definition 2} \textbf{(Adversarial agent).} We call an agent $j$ \textit{adversarial} if under some adversarial control action $u_j^a(t)$, {either of the following holds:}  
\begin{enumerate}
    \item {It performs chasing to hit an intact agent $i$ in some finite time, i.e.,  $\exists \,\,t_a<\infty$ such that $\left\| p_{i}(t) -p_{j}(t) \right\| \to 0$ as $t\rightarrow t_a$.} 

     \item  {It aims to mislead agents in the set ${{\cal V}_f}$ so that they converge to a location $\bar{p}(t) = \frac{1}{{\left| {{{\cal V}_f}} \right|}}\sum\limits_{i \in {{\cal V}_f}} {{p_i}(t)}$ such that $\left\| \bar{p}(t) - G_f \right\| \neq 0$ as $t \to \infty$.}

\end{enumerate}
We denote the overall set of adversarial agents by $\mathcal{A}=\mathcal{A}_s \cup \mathcal{A}_f$, where $\mathcal{A}_s$ and $\mathcal{A}_f$ represent set of adversarial agents correspond to classes 1 and 2 in definition, respectively. We called an agent \textit{intact} if it is not adversarial.
  We denote the set of intact agents as $ \mathcal{V}/\mathcal{A}$.\par

\smallskip

\noindent
\textbf{Definition 3} \textbf{(Proactive adversary detection)}. {We call an adversary detection mechanism {for agent $i$} \textit{proactive}, if detection of the adversary happens at some time $t_d < t_a$ where $t_a$ is given in Definition 2. In particular, adversary detection happens before any adversarial agent $k \in \{i,j\}$  violates the inter-agent safety constraint in \eqref{s1} (i.e., either agent $i$ itself or any neighbor agent $j$ violates the inter-agent safety constraint).}




\smallskip

We define the following as the desired objectives for the intact agent $i \in \mathcal{V}/\mathcal{A}$: 

\begin{enumerate}[{A}.1]
\item  Satisfy constraints on control input, $u_{i}(t) \in {\rm {\mathcal U}}_{i} (x_{i}(t) )=\{ u\in \mathbb{R}^{m_{i} } :A_{i} (x_{i}(t) )u\le b_{i} (x_{i}(t))\}.$

\item  Ensure the forward invariance of the safe set,  $p_{i}(t) \in  S_{i}^{s}$, for all $t>0$.

\item  Guarantee convergence of the closed-loop trajectories to the goal region, i.e., $ V_{i}^{g}(p_i(t)) \to 0$ as $t \to \infty$. 

\item  Maintain a formation over the time interval $t \in [t_a,\,\,t_b]$ such that $\mathop {\lim }\limits_{t \to t_b }  h_i^{{{\cal F}_i}}(\vec p (t)) \to 0,\,\,\forall i \in {{\cal V}_f}$.

\item Based on behavior metrics that capture the degree of satisfaction of the desired objectives A.1-A.4,  proactively determine the set of adversarial agents $\mathcal{A} \subset{\mathcal{V}}$.
\end{enumerate}



{In this work, we assumed that the adversarial agent has the same controller constraints on control input as mentioned in condition A.1.}

Based on the objectives A.1-A.5, the problem formulation is presented as follows.

\smallskip

\noindent 

\noindent \textbf{\textit{Problem 1.}} Consider the desired objectives A.1-A.5. Design control law $u_{i}(t) \in {\rm {\mathcal U}}_{i} (x_{i}(t) )$ for the system \eqref{eq1} such that objectives A.1-A.4 are satisfied for each intact agent $ i\in {\rm {\mathcal V/\mathcal{A.}}}$ 

\vspace{-0.45cm}






\subsection{Forward Invariance of a Set}

 In this subsection, under the assumption of no adversarial agents, i.e., $\mathcal{A}=\emptyset$, we review the necessary and sufficient conditions for guaranteeing forward invariance of a set $S_{i}^{s}$, known also as Nagumo's Theorem. 

\vspace{-0.12cm}
\begin{lemma}
 Let the solution of \eqref{eq1} exist and be unique in forward time. Then, for each agent $i\in {\rm {\mathcal V}}$, the set $S_{i}^{s} $ is forward invariant for the closed-loop trajectories of \eqref{eq1} for all $p_{i}(0) \in S_{i}^{s} $ if and only if the following condition holds:
\begin{equation} \label{ZEqnNum243038} 
{\mathop{\inf }\limits_{u_{i}(t) \in {\rm {\mathcal U_i}}}} \{ {\rm {L}}_{f_{i}^p }h_{i}^{s}(\vec{p})+{\rm {L}}_{g_{i}^p } h_{i}^{s}(\vec{p}) u_{i}(t) \le 0\} ,\, \, \forall p_{i}(t) \in \partial S_{i}^{s},  
\end{equation} 
where $\partial S_{i}^{s}$ represents the boundary of the safe set $S_{i}^{s} .$
\end{lemma}
\vspace{-0.2cm}

{Interested readers can refer to  \cite{blanchini1999set} for more details on forward invariance of sets.}


To ensure the feasibility of Problem 1 when $\mathcal A=\emptyset$, we make the following assumption. In the existing literature, similar assumptions have been used either explicitly or implicitly  (see e.g. \cite{ames2016control}).

\noindent \textbf{Assumption 1.} The trajectories of each agent $i\in {\rm {\mathcal V/\mathcal{A}}}$ satisfy the condition \eqref{ZEqnNum243038}, for all $p_{i} \in \partial S_{i}^{s} \, .$

\noindent \textbf{Assumption 2.} The interior of the set ${\rm {\mathcal U}}_{i} (x_{i}(t) )$ is nonempty and ${\rm {\mathcal U}}_{i} (x_{i}(t) )$ is uniformly compact near $x_{i}(t)$.

\noindent \textbf{Assumption 3.} The functions $f_i$ and $g_i$ are locally Lipschitz with Lipschitz constants $b_f \in \mathbb{R_{+}}$ and $b_g\in \mathbb{R_{+}}$, $\forall i \in \mathcal{V}$, respectively.

\vspace{-0.3cm}

\begin{lemma}
{If the initial} conditions for an agent $i\in {\rm {\mathcal V}}$ are such that $h_{i}^{s}(\vec{p}(0))<0$ and the inequality
\begin{equation} \label{7)} 
{\mathop{\inf }\limits_{u_{i}(t) \in {\rm {\mathcal U}}_{i} }} \{ {\rm {L}}_{f_{i}^p } h_{i}^{S} (\vec{p}(t))+{\rm {L}}_{g_{i}^p } h_{i}^{S} (\vec{p} (t))u_{i}(t) \le \alpha (-h_{i}^{S} (\vec{p}(t)))\} ,\, \,  
\end{equation} 
holds for some locally Lipschitz class-${\rm {\mathcal K}}$ function $\alpha $ for all $t\ge 0,$ then for any $T>0,\, \, \, h_{i}^{s} (x_{i} (t))<0,\, \, \, \forall \, 0\le t\le T.$ 
\end{lemma}

\vspace{-0.45cm}

\subsection{A Quadratic Program for Safety-Control Synthesis}

This subsection presents a quadratic program (QP) to compute a control input $u_{i}(t) $ for each agent $i \in \cal{V}$ to solve Problem 1 when $\mathcal A=\emptyset$. Let $\vec z= [z_1^T,z_2^T,\dots,z_N^T]^T$ be a column vector with  $z_{i} =[u_{i} ,\delta _{i_{1} } ,\delta _{i_{2} },\delta _{i_{3} }\in \mathbb{R}^{m_{i} + 3}],\,\, \forall i\in \cal{V}$ as its elements. Consider the following optimization problem
\vspace{-0.15cm}
\begin{subequations}\label{eq4}
\begin{align}
\begin{split}\label{eq4a} 
 \mathop {\min \,}\limits_{{u_i},{\delta_{{i_1}}},{\delta _{{i_2}}},{\delta _{{i_3}}},\,\,i \in {\cal V}} \,\,\,{{\vec z}^T}H\vec z + F\vec z 
 \vspace{-0.1cm}
\end{split}\\ 
\begin{split} \label{eq4b}
s.t.\,\,\,\,\,\,{A_i}{u_i} \le {b_i},
\end{split}\\
\begin{split}\label{eq4c}
{L_{{f_i^p}}}V_{i}^{g}+ {L_{{g_i^p}}}V_{i}^{g}{u_i} \le  - {\delta _{{i_1}}}V_{i}^{g},\,\,\,\,\forall i \in \mathcal{V}/\mathcal{V}_f,\\ 
\end{split}\\
\begin{split}\label{eq4d}
{L_{{f_i^p}}}h_i^{{{\cal F}_i}} + {L_{{g_i^p}}}h_i^{{{\cal F}_i}}{u_i} \le  - {\delta _{{i_2}}}h_i^{{{\cal F}_i}},\,\,\,\,\forall i \in {{\cal V}_f},\\ 
\end{split}\\
\begin{split}\label{eq4e}
{L_{{f_i^p}}}h_i^S + {L_{{g_i^p}}}h_i^S{u_i} \le  - {\delta _{{i_3}}}h_i^S,\\ 
\end{split}
\end{align}
\vspace{-0.6cm}
\end{subequations}

\noindent
where $H=diag\{H_i\}$ with $H_{i} =diag\{ \{ w_{u_{l} }^{i} \} ,w_{1}^{i} ,w_{2}^{i},w_{3}^{i} \}$ denotes  a diagonal matrix with positive weights $w_{u_{l} }^{i} ,w_{1}^{i} ,w_{2}^{i},w_{3}^{i} >0$  and similarly, $F=diag\{F_i\}$  with $\,\,F_{i} =[0_{m_{i} }^{T}\, \, \, q^{i} \, \,\, 0\, \,\,0]$ where $q^{i} >0$ and $0_{k} \in \mathbb{R}^{k}$ denotes a column vector consisting of zeros. Control input constraints are encoded in \eqref{eq4b}. Performance-based constraints, i.e., goal reaching constraints and formation are encoded in \eqref{eq4c} and \eqref{eq4d}. Similarly, safety-based constraints are encoded in \eqref{eq4e} (which can include both inter-agent safety constraints and obstacle avoidance constraints in conjunction form, as defined in \eqref{eq0}).  {The control formulation in (14) encodes the goal reaching and the safety constraints in (14c) and (14e). Thus, if the QP  is feasible, then the trajectories of the system remain in the safe set and evolve downwards the level sets of the $V_i^g$ towards the goal locations.} 

\vspace{-0.35cm}

\section{Behavior Analysis and Detection Mechanism}

\vspace{-0.1cm}
\subsection{Preliminaries}
{In this subsection, inspired by worst-case designs in the CBF literature \cite{prajna2007framework,taylor2020control,usevitch2021adversarial}, we first introduce the notion of the best and worst-case control actions for the intact and adversarial agent, respectively.} Then, we present the concepts of the critical-time period and critical zone for the agent's operation, which are later required for proactive behavior monitoring, adversary detection and mitigation. In order to provide the best safety guarantee, the minimum pointwise control action by an intact agent $i \in \mathcal{V}/\mathcal{A}$ is defined as
\vspace{-0.15cm}
\begin{align}\setcounter{equation}{14}
\begin{gathered}
u_i^{\min }(t) = \mathop {\arg \min }\limits_{{u_i} \in {{\cal U}_i}} \,\,{L_{{f_i^p}}}h_i^S(\vec p) + {L_{{g_i^p}}}h_i^S(\vec p){u_i},
\end{gathered} 
\label{eq8a}
\vspace{-0.15cm}
\end{align}
where $h_i^S(\vec p)$ in \eqref{eq0} encodes all pairwise safety constraints, i.e., inter-agent and agent-to-obstacle safety constraints. $u_i^{\min }(t)$ in \eqref{eq8a} can be determined by solving the linear program (LP)
\vspace{-0.25cm}
\begin{align}
\begin{gathered}
\mathop {\min }\limits_{{u_i} \in {{\cal U}_i}} \,\,m(\vec p)^T{u_i},\,\,\,\,\,\,
s.t.\,\,\,\,\,\,{A_i}{u_i} \le {b_i},
\end{gathered} 
\label{eq8b}
\vspace{-0.25cm}
\end{align}
where $m(\vec p)^T={L_{{g_i^p}}}h_i^S(\vec p)$. {Note that the pointwise minimum control input provides the best action towards maintaining safety, as the LP in (16) determines the pointwise minimum control action that accounts only for the safety constraints $h_i^S(\vec p)$, without considering any goal reaching objective as encoded in (14c) in terms of $V_i^g$. Thus, $u_i^{min}(t)$ in (15) acts as a safety shield/filter. It is leveraged for the adversary detection and the resilient control design of the multi-agent systems in Sections IV.B, IV.C and V.} 

Similarly, to achieve worst-case safety behavior, the maximum pointwise control action by an adversarial agent $j \in \mathcal{A}$ is defined as
\vspace{-0.2cm}
\begin{align}
\begin{gathered}
u_j^{\max }(t) = \mathop {\arg \max }\limits_{{u_j} \in {{\cal U}_j}} \,\,{L_{{f_j^p}}\bar h_j^s(\vec {p}) + {L_{{g_j^p}}}\bar h_j^s(\vec {p}){u_j}},
\end{gathered} 
\label{eq8c}
\vspace{-0.3cm}
\end{align}
{where $\bar h_j^s(\vec {p})$ in \eqref{s1} encodes only agent-to-agent safety constraints. {In similar fashion, $u_j^{\max }(t)$ in \eqref{eq8c} can be determined by solving the following LP
\vspace{-0.2cm}
\begin{align}
\begin{gathered}
\mathop {\max }\limits_{{u_j} \in {{\cal U}_j}} \,\,m(\vec p_j)^T{u_j},\,\,\,\,\,\,
s.t.\,\,\,\,\,\,{A_j}{u_j} \le {b_j},
\end{gathered} 
\label{eq8d}
\vspace{-0.3cm}
\end{align}
with $m({p_j})^T={L_{{g_j^p}}}\hat{h}_j^s({p_j})$.} Note that in order to achieve desired adversarial chasing behavior as defined in Definition 2, adversarial agent $j$ only needs to account for inter-agent safety. $u_j^{\max}(t)$ denotes the worst control effort by an adversarial agent $j \in \mathcal{A}$ in order to maximize ${L_{{f_j^p}}}\hat{h}_j^s({p_j}) + {L_{{g_j^p}}}\hat{h}_j^s({p_j}){u_j}$ and achieve worst-case safety (i.e., best effort to achieve unsafe behavior).  The feasibility of LP in \eqref{eq8b} is guaranteed under Assumption 2, see \cite{usevitch2021adversarial}.} 
 
 
\vspace{-0.4cm} 
\subsection{Critical Time and Critical Zone}
 \vspace{-0.12cm}
{{Now, in order to design behavior monitors and an adversary detection mechanism, we first present the concept of the critical time period and critical zone for the system \eqref{eq1} in this subsection. We define the critical time period $T_s$, and compute $T_s$ based on the inter-agent safety constraints set $\bar S_i^s$ in \eqref{s2}.} We define inter-agent distance as} 
\begin{equation}\label{c3}
{r_{ij}}({t}) = \left\| {{p_i}({t}) - {p_j}({t})} \right\|, \,\, \forall t>0.
\end{equation}

\noindent \textbf{Definition 4.} 
{$T_s$ is the \textbf{critical time period} of system \eqref{eq1} at the time $t_k$, if under the best-case control input \eqref{eq8a} computed at the time $t_k$, i.e., $u_{i}(t)=u_{i}^{min}({t_k}), \,\, \forall t \in [t_k,\, t_k+T_s]$,  $p_i(t_k) \in int(\bar{S}_{i}^{s})$ implies that $p_i(t_k+T_s) \in \partial \bar{S}_{i}^{s}$.}


\smallskip
\begin{theorem}
Let Assumption $3$ hold. Consider the agent dynamics \eqref{eq1} along with the best-case control input $u_{i}^{min}({t_k})$  at the time $t_k$ given by \eqref{eq8a}. Then, for the safe set $ \bar S_i^s$ in \eqref{s2},  the critical time period is given by $T_s=\mathop {\min \,}\limits_{j \in \mathcal{N}_i} \{T_s^j\}$ with \vspace{-0.1cm}
\begin{equation}
{T_s^j} \ge \frac{1}{{({b_f} + {b_g}\left\| {u_i^{\min }({t_k})} \right\|)}}\log (\frac{1}{{1 -  {\frac{{{r_{ij}}({t_k}) - d}}{{k_1({t_k})}}} }}),
\label{eq_D0}
\vspace{-0.1cm}
\end{equation}
where ${r_{ij}}({t_k})$ is defined in \eqref{c3}, ${b_f}$ and ${b_g}$ denote the Lipschitz constant for functions $f_i(x_i(t))$ and $g_i(x_i(t))$ in \eqref{eq1}, $d$ represents the inter-agent safety distance and
\vspace{-0.1cm}
\begin{equation} 
k_1(t_k)={r_{ij}}({t_k})+\frac{{b_g}\left\| {p_j}(t_k)  \right\| \left\| u_{j}^{max}({t_k})-u_{i}^{min}({t_k}) \right\|}{{b_f} + {b_g}\left\| {u_i^{\min }({t_k})} \right\|}.
\label{eq_D001}
\vspace{-0.1cm}
\end{equation}

\end{theorem}

\begin{proof}
{Define the variation of the inter-agent safety constraint \eqref{s1} over the time interval $[t_k, t_k+t]$ as}
\vspace{-0.15cm}
\begin{equation}
\Upsilon (t + {t_k},{t_k}) = h_{ij}^{s} (p_{i}(t+t_k),\,p_{j}(t+t_k) ) - h_{ij}^{s} (p_{i}(t_k),\,p_{j}(t_k) ),
\label{eq_D1}
\vspace{-0.15cm}
\end{equation}
which can be written as
\vspace{-0.25cm}
\begin{equation}
\begin{array}{l}
\Upsilon (t + {t_k},{t_k})
 ={r_{ij}}({t_k}) - {r_{ij}}(t+{t_k}),
\end{array}
\label{eq_D2}
\vspace{-0.1cm}
\end{equation}
and its derivative can be computed as 
\vspace{-0.2cm}
\begin{equation}
\begin{array}{l}
\dot \Upsilon (t + {t_k},{t_k}) =  - \frac{{{{({p_i}(t + {t_k}) - {p_j}(t + {t_k}))}^T}}}{{\left\| {{p_i}(t + {t_k}) - {p_j}(t + {t_k})} \right\|}}({{\dot p}_i}(t + {t_k}) - {{\dot p}_j}(t + {t_k})),
\end{array}
\label{eq_D3}
\vspace{-0.1cm}
\end{equation}
where the term $\frac{{{{({p_i}(t + {t_k}) - {p_j}(t + {t_k}))}^T}}}{{\left\| {{p_i}(t + {t_k}) - {p_j}(t + {t_k})} \right\|}}$ is a unit vector and thus, 
\begin{equation}
\dot \Upsilon (t + {t_k},{t_k})  \le  \left\|({{{\dot p}_j}(t + {t_k}) - {{\dot p}_i}(t + {t_k})})\right\|.
\label{eq_D4}
\end{equation}
From the triangular inequality, one has
\vspace{-0.2cm}
\begin{equation}
\begin{array}{l}
 {\dot \Upsilon (t + {t_k},{t_k})}  \le \left\| {f_j^p({x_j}(t + {t_k})) - f_i^p({x_i}(t + {t_k}))} \right\| \\ \,\,\,\,\,\,\,\,\, \,\,\,\,\,\,\,\,\,\,\,\,\,\,\,\,\,\, \,\,\,\,\,\,\,\,\,\,\,\, + \left\| {g_j^p({x_j}(t + {t_k})){u_j}(t + {t_k}) - g_i^p({x_i}(t + {t_k})){u_i}(t + {t_k})} \right\|,\\
\end{array}
\label{eq_D5}
\end{equation}
where $f_i^p \in \mathbb{R}^{2}$ and $g_i^p \in \mathbb{R}^{{2} \times {m_i}}$ denotes the sub-matrix of $f_i$ and $g_i$ corresponding to position vector dynamics in \eqref{eq1}. Based on Assumption $3$, $f_i^p$ and $g_i^p$, $\forall i \in \mathcal{V}$, are locally Lipschitz and are bounded by the Lipschitz constants $b_f \in \mathbb{R_{+}}$ and $b_g\in \mathbb{R_{+}}$, respectively. Thus, under the constant best control input evaluated at the time $t_k$, i.e., $u_{i}(t)=u_{i}^{min}({t_k}), \,\, \forall t \in [t_k,\, t_k+T_s^j]$, equation \eqref{eq_D5} becomes
\begin{equation}
\begin{array}{l}
 {\dot \Upsilon (t + {t_k},{t_k})}  \le ({b_f} + {b_g}\left\| {u_i^{\min }({t_k})} \right\|)\left\| {({p_i}(t + {t_k})) - ({p_j}(t + {t_k}))} \right\|\\\,\,\,\,\,\,\,\,\, \,\,\,\,\,\,\,\,\,\,\,\,\,\,\,\,\,\, \,\,\,\,\,\,\,\,\,\,\,\,\,\,\,\,\,\,\,\,\,\,+\Delta ({p_j}(t_k)),
\end{array}
\label{eq_D6}
\end{equation}
with $\Delta ({p_j}(t_k))= {b_g}\left\| {p_j}( {t_k})  \right\| \left\| u_{j}^{max}({t_k})-u_{i}^{min}({t_k}) \right\|$. Now based on the defined error term $\Upsilon(t + {t_k},{t_k})$ in  \eqref{eq_D1}, one can write
\begin{equation}
\begin{array}{l}
 {\dot \Upsilon (t + {t_k},{t_k})} \le  - ({b_f} + {b_g}\left\| {u_i^{\min }({t_k})} \right\|)\Upsilon (t + {t_k},{t_k}) \\ \,\,\,\,\,\,\,\,\, \,\,\,\,\,\,\,\,\,\,\,\,\,\,\,\,\,\, \,\,\,\,\,\,\,\,\,\,\,\,\,\,\,\,\,\,\,\,\,\,\,\,\, \, \,\,\,\, + ({b_f} + {b_g}\left\| {u_i^{\min }({t_k})} \right\|){r_{ij}}({t_k})+ \Delta ({p_j}(t_k)).\\
\end{array}
\label{eq_D7}
\end{equation}
Then, based on Comparison Lemma \cite{khalilNL}, one has following solution
\vspace{-0.2cm}
\begin{equation}
 {\Upsilon (t + {t_k},{t_k})}  \le k_1(t_k)(1 - {e^{ - ({b_f} + {b_g}\left\| {u_i^{\min }({t_k})} \right\|)(t - {t_k})}}),
\label{eq_D8}
\end{equation}
with $k_1(t_k)$ in \eqref{eq_D001}. Now, with $h_{ij}^{s} (p_{i},\,p_{j} )$ defined in \eqref{s1}, we know that at the time instant $t=T_s^j$ from \eqref{eq_D1}, one has $ {\Upsilon (t_k+T_s^j,{t_k})}  =  {{r_{ij}}({t_k}) - d}$ as $h_{ij}^{s} (p_{i}(t_k+T_s^j),\,p_{j}(t_k+T_s^j))$ becomes zero as $p_i(t_k+T_s^j) \in \partial \bar{S}_{i}^{s}$, i.e., $ d-{{r_{ij}}({t_k+T_s^j}) = 0}$. Then based on \eqref{eq_D8}, the bound on critical time period $T_s^j$ can be computed as
\vspace{-0.2cm}
\begin{equation}
{\frac{{{r_{ij}}({t_k}) - d}}{k_1(t_k)}} \le (1 - {e^{ - ({b_f} + {b_g}\left\| {u_i^{\min }({t_k})} \right\|){T_s^j}}}),
\label{eq_D9}
\end{equation}
which finally yields \eqref{eq_D0}. This completes the proof.
\end{proof}

\vspace{-0.3cm}

In the following theorem, we present the result for evaluation of the critical time period $T_s^{o}$ for the conjunction of agent-to-obstacle safe set $ \bar S_i^s$ in \eqref{s4}.


\vspace{-0.1cm}
\begin{theorem}
Let Assumption $3$ hold. Consider the agent dynamics \eqref{eq1} along with the worst-case control input in \eqref{eq8c} evaluated at the time $t_k$. Then, for the safe set $ \hat S_i^s$ in \eqref{s4}, the critical time period $T_s^{o}=\mathop {\min \,}\limits_{{oj} \in \mathcal{O}_{i}} \{T_s^{oj}\}$ with
\vspace{-0.2cm}
\begin{equation}
{T_s^{oj}} \ge \frac{1}{{({b_f} + {b_g}\left\| {u_i^{\max }({t_k})} \right\|)}}\log (\frac{1}{{1 -  {\frac{{{v_{i}^{oj}}({t_k}) - r_{oj}}}{{{v_{i}^{oj}}({t_k})+r_{oj}}}} }}),
\label{eq_D01}
\vspace{-0.2cm}
\end{equation}
where ${v_{i}^{oj}}({t_k}) = \left\| {{p_i}({t_k}) - c_{oj}} \right\|$ and, $c_{o_j}$ and $r_{o_j}$ are defined in \eqref{s3}. Moreover, ${b_f}$ and ${b_g}$ denote the Lipschitz constant for functions $f_i(x_i(t))$ and $g_i(x_i(t))$ in \eqref{eq1}. 

\end{theorem} 

\begin{proof}
The result follows a similar argument as given in the proof of Theorem 1 with agent-obstacle pairwise safety function  $h_i^{o_j} (p_{i}(t))$ in \eqref{s3} instead of inter-agent safety function $h_{ij}^{s}(p_{i},\,p_{j} )$ in \eqref{s1}. 
\end{proof}

\noindent \textbf{Definition 5.} The critical zone $\eta(p_i(t), p_j(t)) : \mathbb{R}^{2} \to \mathbb{R}$ is defined as 
\vspace{-0.2cm}
\begin{align}
\begin{gathered}
 \eta(p_i(t), p_j(t))= \mathop {\max \,}\limits_{u_i(t),u_j(t)} \left\|  \int\limits_{t}^{t+nT_s} (\dot{p}_i(\tau) - \dot{p}_j(\tau))d\tau  \right\|,\\
\end{gathered} 
\label{eq13de}
\vspace{-0.2cm}
\end{align}
which represents the maximum magnitude of the evolution of the difference between the position trajectories of an intact agent $i$ and its neighbor $j$ over the time interval $[t,\,t+nT_s]$, with $n \in \mathbb{Z}_{>1}$ and $T_s$ being the critical time period provided in Definition $4$. 

\vspace{-0.15cm}

\begin{remark}
Note that  $nT_s$ in Definition $5$ denotes the desired sampling time for trajectory evaluation of agent $i$ and it can be designed such that future safety is always ensured. Based on Theorem 1, under the best-case control input, i.e., $u_{i}^{min}(t_k)$, the agent $i$ reaches the  boundary of the safe set $ \bar S_i^s$ in \eqref{s2} over the time interval $[t_k,\,t_k+T_s]$. That's why the critical region $\eta(p_i(t), p_j(t))$ is evaluated over horizon $[t,\,t+nT_s]$ with the design parameter $n \in \mathbb{Z}_{>1}$ and augmented with the inter-agent safety constraints such that it that provides robustness to actual safe region. Then, we leverage this notion to proactively detect adversarial agent without violating safety constraints and ensure all time safety for all $i \in \mathcal{V}/\mathcal{A}$ based on resiliency mechanism as presented in Section IV. C and V, respectively. 
\end{remark} 
\vspace{-0.15cm}

Note that $\eta(p_i(t), p_j(t))$ can be maximized by applying best and worst-case control input $u_{i}^{min}({t})$ in \eqref{eq8a} and $u_{j}^{max}({t})$ in \eqref{eq8c}, respectively, over the time interval $[t,\,t+nT_s]$. Thus, based on Definition $5$, we can rewrite critical zone as 
\vspace{-0.1cm}
\begin{align}
\begin{gathered}
 \eta(p_i(t), p_j(t))=   \,\,\,\,\,\, \,\,\,\,\,\, \,\,\,\,\,\, \,\,\,\,\,\,\,\,\,\,\,\, \,\,\,\,\,\, \,\,\,\,\,\, \,\,\,\,\,\,\\ \left\| \int\limits_{t}^{t+t_s} (f_{i}^p (\tau)+g_{i}^p ( \tau)u_{i}^{min}-f_{j}^p (\tau)-g_{j}^p ( \tau)u_{j}^{max})d\tau   \right\|.
\end{gathered} 
\label{eq13e}
\vspace{-0.1cm}
\end{align}

\noindent
where $f_i^p \in \mathbb{R}^{2}$ and $g_i^p \in \mathbb{R}^{{2} \times {m_i}}$ denotes the sub-matrix of $f_i$ and $g_i$ corresponding to position vector dynamics in \eqref{eq1}.  For brevity, we denote $\eta(p_{ij}(t))=\eta(p_i(t), p_j(t))$ in the rest of the paper. Also, the presented formulation for the critical zone $\eta(p_{ij}(t))$ can be directly extended for the static obstacle avoidance case with obstacle at fixed position, i.e.,
\vspace{-0.1cm}
\begin{align}
\begin{gathered}
 \eta(p_{i_{o_j}}(t))= \left\| \int\limits_{t}^{t+\bar{t}_s} (f_{i}^p (\tau)+g_{i}^p (\tau)u_{i}^{max} )d\tau   \right\|, \\
\end{gathered} 
\label{eq13def}
\end{align}
\vspace{-0.1cm}
\noindent
where $\bar{t}_s=nT_s^{oj}$ with $T_s^{oj}$ defined in \eqref{eq_D01}. The designed critical zones $\eta(p_{ij}(t))$ and $\eta(p_{i_{o_j}}(t))$ are leveraged in Section IV.B for behavior monitoring and design of proactive adversary detection mechanism.  

\vspace{-0.4cm}

\subsection{Behavior Metrics for Monitoring and Proactive Adversary Detection}

In this subsection, we first introduce behavior metrics and then design behavior monitors for the detection of adversarial agents belonging to the set $ \mathcal{A}_s$. We present the following behavior metrics: (i) Safety behavior metric, and (ii) Goal reaching behavior metric.


\noindent
{\textit{ Safety behavior metric:} To monitor the agents behavior for safe operation, the \textbf{safety behavior metric} is defined as
\vspace{-0.2cm}
\begin{equation}\label{eq10}
    S_{{R_i}}(t) = \exp {( - (\Gamma _i(p_i(t),p_j(t)))^{{n_c}}}),\,\,\forall t \geq 0,
\end{equation}
where
\vspace{-0.25cm}
\begin{equation}\label{eq11}
\Gamma_i(p_i(t),p_j(t))= \max \{ {g_{ij}({p_i(t)},{p_j(t)}) },{g_{io_{j}}({p_i(t)}) }\},
\end{equation}
with
\vspace{-0.33cm}
\begin{equation}\label{eq11a}
{g_{ij}}({p_i(t)},{p_j(t)}) = \frac{d}{{{{\left\| {{p_i(t)} - {p_j(t)}} \right\|}}}},\,\,\forall {j} \in {\mathcal{N}_i},
\end{equation} and 
\vspace{-0.2cm}
\begin{equation}\label{eq11b}
{g_{i{o_j}}}({p_i(t)}) = \frac{{r_{{o_j}}}}{{{{\left\| {{p_i(t)} - {c_{{o_j}}}} \right\|}}}},\,\,\forall {o_j} \in {\mathcal{O}_i},
\end{equation} 
where $c_{o_j}$ and $r_{o_j}$ $c_{o_j}$ and $r_{o_j}$ are defined in \eqref{s3}. Moreover, ${\mathcal{O}_i}$ and $n_c \in \mathbb{Z}_{>1}$ denote the set of obstacles for the agent $i$ and a constant design gain, respectively. Now the following proposition shows how the designed safety behavior metric $S_{{R_i}}(t)$  acts as a safety monitor for an agent $i$.}
\vspace{-0.15cm}
\begin{proposition}
{Consider the agent dynamics \eqref{eq1} along with the safety behavior metric $S_{{R_i}}(t)$ in \eqref{eq10}. For the defined safe set $S_{i}^{s}$ in \eqref{s5}, if $p_i(t) \in int(S_{i}^{s})$ ($p_i(t) \notin S_{i}^{s}$), then the safety behavior metric $S_{{R_i}}(t) \to 1$ ($S_{{R_i}}(t) \to 0$) for all time $t$.}
\end{proposition}

\begin{proof}
{Note that, based on safety constraints in \eqref{s1} and \eqref{s3}, for  any $p_i(t) \in int(S_{i}^{s})$,  one has  ${g_{ij}}({p_i(t)},{p_j(t)})  < 1$ and ${g_{i{o_j}}(p_i(t))}<1$. Similarly, for  any $p_i(t) \notin S_{i}^{s}$,  one has  ${g_{ij}}({p_i(t)},{p_j(t)}) > 1$ and ${g_{i{o_j}}(p_i(t))}> 1$. Thus, based on \eqref{eq11a}, \eqref{eq11b} and \eqref{eq10}, the safety behavior metric $S_{{R_i}}(t) \to 1$ ($S_{{R_i}}(t) \to 0$) if $p_i(t) \in int(S_{i}^{s})$ ($p_i(t) \notin S_{i}^{s}$) with a proper design constant $n_c \in \mathbb{Z}_{>1}$. Therefore, based on the metric $S_{{R_i}}(t), \,\forall t \geq 0,$ one can employ $S_{{R_i}}(t)$ as a monitor to evaluate the safe (unsafe) behavior of the agent. }
\end{proof}


\vspace{-0.2cm}
{Now, we leverage the presented critical zone in Section IV. B and define a metric to capture the worst-case safety behavior in terms of critical zones as}
\vspace{-0.2cm}
\begin{equation}\label{eq13a}
    S_{{R_i}}^w(t) = \exp {( - \Gamma_{i}^{w}(p_i(t),p_j(t))^{{n_c}}}),\,\,\forall t \geq 0,
    \vspace{-0.2cm}
\end{equation}
{where}
\vspace{-0.2cm}
\begin{equation}
\Gamma_{i}^{w}(p_i(t),p_j(t)) = 1- \max \{ {g_{ij}^w(p_i,p_j) }, \, {g_{io_{j}}^w(p_i) }\}, 
\label{eq13b}
\end{equation} 
with
\vspace{-0.2cm}
\begin{equation}\label{eq13c}
g_{ij}^w (p_i(t),p_j(t))= \frac{\eta(p_{ij}(t))}{{\left\| {{p_i(t)} - {p_j(t)}} \right\|}},
\end{equation} and 
\vspace{-0.2cm}
\begin{equation}\label{eq13d}
{g_{i{o_j}}^w}({p_i(t)}) =  \frac{\eta(p_{i{{o_j}}}(t))}{{\left\| {{p_i(t)} - {c_{{o_j}}}} \right\|}},
\end{equation}
$\forall i \in {\rm {\mathcal V}} ,\,\,{o_j} \in {\mathcal{O}_i},$ with $\eta(p_{ij}(t))$ and $\eta(p_{i{{o_j}}}(t))$ critical zones defined in \eqref{eq13de} and \eqref{eq13def}, respectively. We define the error between the worst-case and nominal safety metrics as
\begin{equation}\label{c1}
\gamma_i^S(t) = \left\| {S_{{R_i}}^w(t) - S_{{R_i}}^n(t)} \right\|,
\end{equation}
with ${S_{{R_i}}^n(t)}=1$ as the nominal safety behavior metric.  Note that ${S_{{R_i}}^n(t)}$ represents safety behavior metric $S_{{R_i}}(t)$ in \eqref{eq10} under nominal operation of agent $i$, i.e., when  $p_i(t) \in int(S_{i}^{s})$. 

{Now, we present the first result on behavior monitoring, and show that if the safety behavior metric value is below a designed threshold, then one can always ensure the operation of an agent inside the safe set $S_{i}^{s}$.}

\vspace{-0.2cm}

\smallskip
\begin{theorem}
Consider the agent dynamics \eqref{eq1} along with the QP based control  \eqref{eq4} and the safety behavior metric $S_{{R_i}}(t)$ in \eqref{eq10}. If it holds that 
\vspace{-0.2cm}
\begin{equation}
{\left\| {{S_{{R_i}}}(t) - S_{{R_i}}^n(t)} \right\|} \leq {\gamma _i^S(t)},\,\,\forall t \geq 0
\label{m7}
\end{equation}
where $\gamma_i^S(t)$ defined is \eqref{c1}, 
then $p_i(t) \in int(S_{i}^{s}),\,\, \forall t \geq 0$. 
\end{theorem}


\begin{proof}
If
\vspace{-0.25cm}
\begin{equation}
{\left\| {{S_{{R_i}}}(t) - S_{{R_i}}^n(t)} \right\|} \leq \left\| {S_{{R_i}}^w(t) - S_{{R_i}}^n(t)} \right\|,
\label{eq14}
\end{equation}
then due to the nominal safety behavior metric ${S_{{R_i}}^n(t)} = 1$, one has $S_{{R_i}}^w(t) \leq S_{{R_i}}(t)$ as $0 \leq S_{{R_i}}^w(t), S_{{R_i}}(t) \leq 1$. From  $S_{{R_i}}(t)$ and $S_{{R_i}}^w(t)$ in \eqref{eq10} and \eqref{eq13a}, one has 
\begin{equation}
\Gamma _i(p_i(t),p_j(t),u_i(t))) \leq \Gamma_{i}^{w}(p_i(t),p_j(t),u_i(t))),
\label{eq15}
\end{equation}
which using \eqref{eq11} and \eqref{eq13b} can be written as
\vspace{-0.2cm}
\begin{equation}
\max \{ {g_{ij}(p_i,p_j) }, \, {g_{io_{j}}(p_i)}\} \leq 1- \max \{ {g_{ij}^w(p_i,p_j) }, \, {g_{io_{j}}^w(p_i) }\}, \vspace{-0.2cm}
\label{eq16a}
\end{equation}
$\forall j \in \mathcal{N}_i$ and $\forall o_{j} \in \mathcal{O}_i,$ and it further  becomes
\vspace{-0.2cm}
\begin{equation}
\max \{ {g_{ij}(p_i,p_j) }+{g_{ij}^w(p_i,p_j)},\, {g_{io_{j}}(p_i)}+{g_{io_{j}}^w(p_i) }\} \leq 1,
\label{eq16b}
\vspace{-0.2cm}
\end{equation}
as ${g_{ij}(p_i,p_j)}$, ${g_{ij}^w(p_i,p_j)}$, ${g_{io_{j}}(p_i)}$ and ${g_{io_{j}}^w(p_i) } \in \mathbb{R}_{>0}$ according to \eqref{eq11a}-\eqref{eq11b} and \eqref{eq13c}-\eqref{eq13d}. Then, based on ${g_{ij}(p_i,p_j)}$, ${g_{ij}^w(p_i,p_j)}$, ${g_{io_{j}}(p_i)}$ and ${g_{io_{j}}^w(p_i)}$, one can write \eqref{eq16b} as
\vspace{-0.2cm}
\begin{equation}
\max \{ \frac{d + \eta ({p_{ij}}(t))}{\left\| {{p_i}(t) - {p_j}(t)} \right\|},\,\frac{{r_{{o_j}}} + \eta ({p_{i{o_j}}}(t))}{\left\| {{p_i}(t) - {c_{{o_j}}}} \right\|} \} \leq 1.
\label{eq16b1}
\vspace{-0.15cm}
\end{equation}
Now, from \eqref{eq16b1}, one can infer that 
\vspace{-0.2cm}
\begin{equation}
(d+\eta(p_{ij}(t)))-{{\left\| {{p_i(t)} - {p_j(t)}} \right\|}} \leq 0, \,\,\forall j \in \mathcal{N}_i,
\label{eq16c}
\vspace{-0.15cm}
\end{equation}
and 
\vspace{-0.3cm}
\begin{equation}
(r_{{o_j}}+\eta(p_{io_{j}}(t)))-{{\left\| {{p_i(t)} - c_{{o_j}}} \right\|}} \leq 0, \,\, \forall o_{j} \in \mathcal{O}_i,
\label{eq16d}
\vspace{-0.2cm}
\end{equation}
as $\eta(p_{ij}(t))$ and $\eta(p_{io_{j}}(t)) \in \mathbb{R}_{>0}$ based on its definition in \eqref{eq13de} and \eqref{eq13def}, respectively. Then, based on \eqref{s5}, it implies $h_i^S(\vec{p}) < 0$ and thus,  $p_i(t) \in int(S_{i}^{s}),\,\, \forall t \geq 0$.
\end{proof}


\vspace{-0.3cm}

\begin{remark}
{Note that based on Theorem 3, we know that if the safety behavior metric value is below a designed threshold, then one can always ensure the operation of an agent inside the safe set $S_{i}^{s}$. However, if the safety behavior metric violates the designed threshold, then one needs to determine whether the safety of the agent is jeopardized due to the agent's own behavior, or due to a neighbor's behavior. Thus, one needs a metric to differentiate the adversarial agent trying to achieve the class 1 objective in Definition 2. Therefore, to accomplish proactive adversary identification in terms of critical time and critical zone, we also need to monitor goal reaching behavior to differentiate any adversarial agent that belongs to class 1 type of adversary and tries to violate safety constraints. Based on the provided reasoning, we design the proactive adversary detection mechanism in Theorem 4.}
\end{remark}

\vspace{-0.2cm}

\noindent
\textit{Goal reaching behavior metric:} {To monitor the agent's performance in terms of reaching toward the goal location, we define the \textbf{goal reaching behavior metric} as}
\vspace{-0.2cm}
\begin{equation}\label{eq12}
{G_{{R_i}}(t)} = \exp {( - ({\lambda _i}({p_i(t)},{p_i}(0),{G_i}))^{{n_c}}}),\,\,\forall t \geq 0,
\vspace{-0.2cm}
\end{equation}
with 
\vspace{-0.3cm}
\begin{align}\label{eq12a}
{\lambda _i}({p_i(t)},{p_i}(0),{G_i}) = \frac{{{{\left\| {{p_i}(t) - {G_i}} \right\|}^2}}}{{{{\left\| {{p_i}(0) - {G_i}} \right\|}^2}}} 
=\zeta V_{i}^{g}(p_i(t)),\,\,\forall t \geq 0,
\vspace{-0.35cm}
\end{align} 
where $\zeta=\frac{1}{{\left\| {{p_i}(0) - {G_i}} \right\|}^2}$, $V_{i}^{g}(p_i(t))$ is defined in \eqref{s6} and ${p_i}(0)$ represents initial position of the agent $i$. Note that $0\leq  \lambda _i({p_i}(t),{p_i}(0),{G_i}) \leq 1$ under normal or desired behavior of the agents.

{In the following proposition, we analyze how the goal reaching metric defined in \eqref{eq12a} changes for an agent when it starts deviating from the desired goal reaching behavior. As explained in Remark 3, we need the result of Proposition 2 along with behavior metrics in Theorem 4 to differentiate any adversarial agent (either agent $i$ or its neighboring agent $j$) that tries to violate safety constraints.}

\vspace{-0.2cm}

\begin{proposition}
Consider the agent dynamics \eqref{eq1} along with the QP based control \eqref{eq4}. For ${\lambda _i}({p_i}(t))$  defined in \eqref{eq12a},  if an agent $i$ satisfies   
\begin{align}
\begin{gathered}
 ({{L}_{f_i^p}{\lambda _i}({p_i}(t)) + {L}_{g_i^p}}{\lambda _i}({p_i}(t)){u_i}(t) \ge 0) \wedge ({\lambda _i}({p_i}(t)) \ne 0),
\end{gathered} 
\label{eq40}
\end{align}
for all $t \in [t_0, t_0+ t_m]$ with some time $t_0\geq 0,\,\,t_m > 0$, then the agent $i$ deviates from desired goal reaching behavior.
\end{proposition}


\vspace{-0.3cm}

\begin{proof}

If condition provided in \eqref{eq40} holds true at some time $t \geq 0$ , then under \eqref{eq12a}, one has
\vspace{-0.1cm}
\begin{equation}\label{m1}
   ({L_{{f_i^p}}}V_{i}^{g}(p_i(t)) + {L_{{g_i^p}}}V_{i}^{g}(p_i(t)){u_i(t)}\ge 0) \wedge (V_{i}^{g}(p_i(t))) \ne 0),
   \vspace{-0.12cm}
\end{equation}
\noindent as $\zeta>0$ in \eqref{eq12a}. That means the agent $i$ has not reached the goal point and it starts deviating from the desired goal reaching behavior over time $t \in [t_0, t_0+ t_m]$ as ${L_{{f_i^p}}}V_{i}^{g}(p_i(t)) + {L_{{g_i^p}}}V_{i}^{g}(p_i(t)){u_i(t)}\ge 0$.  
\end{proof}



\vspace{-0.12cm}

{In the following theorem, we present the proactive adversary detection mechanism based on the designed behavior metrics in \eqref{eq10} and \eqref{eq12} along with the presented critical zone and critical time in Section IV. B.}
\vspace{-0.2cm}

\begin{theorem}
Consider the agent dynamics \eqref{eq1} along with  $S_{{R_i}}(t)$ and $\gamma _i^S(t)$ defined in \eqref{eq10} and \eqref{c1}, respectively. {If an agent $i$ satisfies   
\vspace{-0.2cm}
\begin{equation}
{\rm{}}\left\| {{S_{{R_i}}}(t) - S_{{R_i}}^n(t)} \right\| > \gamma _i^S(t),  \,\,\,\,\,\, \forall t \in [t_0, \,t_0+ nT_s],
\label{eq30}
\vspace{-0.2cm}
\end{equation}
with some time $t_0 \geq 0$ then,}
\begin{enumerate}
\item the { position trajectories of agent $i$ remain in the safe set for a horizon $(n-1)T_s$ ahead}, i.e.,
 $p_i(t) \in int(S_{i}^{s}), \,\, \forall t \in [t_0, t_0+ (n-1)T_s]$.
 \item If in addition $ {{L}_{f_k^p}}{\lambda _k}({p_k}(t)) + {{L}_{g_k^p}}{\lambda _k}({p_k}(t)){u_k}(t) \ge 0$, $\forall t \in [t_0, t_0+ (n-1)T_s]$, where $k\in\{i,j\}$, $j\in \mathcal N_i$, then the agent $i$ detects any agent $k$ (i.e., either itself $i$ or any neighbor agent $j$) as adversarial at time $t=t_0+ (n-1)T_s$.
\end{enumerate}
\end{theorem}


\vspace{-0.2cm}

\begin{proof}
The equation \eqref{eq30} under  
\eqref{eq10} and \eqref{eq13a} with some mathematical simplification, reads
\vspace{-0.15cm}
\begin{equation}
\Gamma _i(p_i(t),p_j(t),u_i(t))) > \Gamma_{i}^{w}(p_i(t),p_j(t),u_i(t))),
\label{eq32}
\end{equation}
which based on \eqref{eq11} and \eqref{eq13b}, becomes
\vspace{-0.15cm}
\begin{equation}
\max \{ {g_{ij}(p_i,p_j) }, \, {g_{io_{j}}(p_i)}\} > 1- \max \{ {g_{ij}^w(p_i,p_j) }, \, {g_{io_{j}}^w(p_i) }\} ,
\label{eq32a}
\vspace{-0.2cm}
\end{equation}
$\forall j \in \mathcal{N}_i$ and $\forall o_{j} \in \mathcal{O}_i$ and thus, one has
\vspace{-0.15cm}
\begin{equation}
\max \{ {g_{ij}(p_i,p_j) }+{g_{ij}^w(p_i,p_j)},\, {g_{io_{j}}(p_i)}+{g_{io_{j}}^w(p_i) }\} > 1,
\label{eq32b}
\vspace{-0.15cm}
\end{equation}
as ${g_{ij}(p_i,p_j)}$, ${g_{ij}^w(p_i,p_j)}$, ${g_{io_{j}}(p_i)}$ and ${g_{io_{j}}^w(p_i) } \in \mathbb{R}_{>0}$ according to \eqref{eq11a}-\eqref{eq11b} and \eqref{eq13c}-\eqref{eq13d}. Then, based on ${g_{ij}(p_i,p_j)}$, ${g_{ij}^w(p_i,p_j)}$, ${g_{io_{j}}(p_i)}$ and ${g_{io_{j}}^w(p_i)}$, one can write \eqref{eq32b} as
\vspace{-0.15cm}
\begin{equation}
\max \{ \frac{d + \eta ({p_{ij}}(t))}{\left\| {{p_i}(t) - {p_j}(t)} \right\|},\,\frac{{r_{{o_j}}} + \eta ({p_{i{o_j}}}(t))}{\left\| {{p_i}(t) - {c_{{o_j}}}} \right\|} \} > 1.
\label{eq32b1}
\end{equation}

Thus, one can conclude at least one of the following conditions do not hold true,  $\forall t \in [t_0, t_0+ (n-1)T_s]$,
\vspace{-0.2cm}
\begin{align}\label{eq32c}
    \left\{ \begin{array}{l}
(d + \eta ({p_{ij}}(t))) - \left\| {{p_i}(t) - {p_j}(t)} \right\| \nleq 0,{\mkern 1mu} {\kern 1pt} {\mkern 1mu} {\kern 1pt} \forall j \in {{\cal N}_i},\\
({r_{{o_j}}} + \eta ({p_{i{o_j}}}(t))) - \left\| {{p_i}(t) - {c_{{o_j}}}} \right\| \nleq 0,{\mkern 1mu} {\kern 1pt} {\mkern 1mu} {\kern 1pt} \forall {o_j} \in {{\cal O}_i},
\end{array} \right.
\vspace{-0.2cm}
\end{align}
with $\eta(p_{ij}(t))$ and $\eta(p_{io_{j}}(t)) \in \mathbb{R}_{>0}$ based on its definition in \eqref{eq13de}. However, the critical zone $\eta(p_{ij}(t))$ and $\eta(p_{io_{j}}(t))$ are designed such that in worst-case scenario the agent can reach unsafe boundary only if $t \geq t_0+ nT_s,$ i.e., when the critical zones shrink to zero. Thus, following safety conditions are satisfied,  $\forall t \in [t_0, t_0+ (n-1)T_s]$,
\vspace{-0.2cm}
\begin{align}\label{eq32d}
    \left\{ \begin{array}{l}
d  - \left\| {{p_i}(t) - {p_j}(t)} \right\| < 0,{\mkern 1mu} {\kern 1pt} {\mkern 1mu} {\kern 1pt} \forall j \in {{\cal N}_i},\\
{r_{{o_j}}} - \left\| {{p_i}(t) - {c_{{o_j}}}} \right\| < 0,{\mkern 1mu} {\kern 1pt} {\mkern 1mu} {\kern 1pt} \forall {o_j} \in {{\cal O}_i},
\end{array} \right.
\vspace{-0.2cm}
\end{align}
This implies $h_i^S(\vec{p}) < 0$ and thus, the safety for agent $i$ is guaranteed, i.e., $p_i(t) \in int(S_{i}^{s}), \,\, \forall t \in [t_0, t_0+ (n-1)T_s]$. This completes the proof of part $1$. 


{Now we prove part 2 of the theorem. Based on the condition in \eqref{eq30} and from the proof of part 1, one can conclude that at least one of the following conditions does not hold true: 
\vspace{-0.2cm}
\begin{align}\label{eq32cu}
    \left\{ \begin{array}{l}
(d + \eta ({p_{ij}}(t))) - \left\| {{p_i}(t) - {p_j}(t)} \right\| \nleq 0,{\mkern 1mu} {\kern 1pt} {\mkern 1mu} {\kern 1pt} \forall j \in {{\cal N}_i},\\
({r_{{o_j}}} + \eta ({p_{i{o_j}}}(t))) - \left\| {{p_i}(t) - {c_{{o_j}}}} \right\| \nleq 0,{\mkern 1mu} {\kern 1pt} {\mkern 1mu} {\kern 1pt} \forall {o_j} \in {{\cal O}_i},
\end{array} \right.
\vspace{-0.0cm}
\end{align}
for all $\forall t \in [t_0, t_0+ nT_s]$ with $\eta(p_{ij}(t))$ and $\eta(p_{io_{j}}(t)) \in \mathbb{R}_{>0}$. Note that if the conditions in \eqref{eq32cu} hold true for any $j \in \mathcal{N}_i$ or $ o_{j} \in \mathcal{O}_i$, $\,\, \forall t \geq t_0+nT_s$, then the safety conditions \eqref{eq32d} do not satisfy $\forall j \in \mathcal{N}_i$ and $ \forall o_{j} \in \mathcal{O}_i$ as the designed critical zone $\eta(p_{ij}(t))$ or $\eta(p_{io_{j}}(t))$ in \eqref{eq13de}  and \eqref{eq13def} becomes zero at $t = t_0+nT_s$. Thus, $p_i(t) \notin int(S_{i}^{s}), \,\, \forall t \geq nT_s$. Therefore, one needs to detect agent $i$ or $j$ as adversarial at $t = t_0+(n-1)T_s$ and act to ensure safety of the intact agent. However, the violation

\vspace{-0.0cm}

\noindent

\begin{algorithm}[!ht]
\caption{Identification of adversarial agents in set $\mathcal{A}_{s}$.}
\begin{enumerate}
\item [1:] Initialize with design constant $n \in $ $\mathbb{Z}_{>1}$.
\item [2:] \textbf{procedure} $\forall i \in \mathcal{V}/\mathcal{V}_f$ 
\item [3:] At each time $t_0$, compute the critical time $T_s=\mathop {\min \,}\limits_{j \in \mathcal{N}_i} \{T_s^j\}$ where $T_s^j$ is defined in \eqref{eq_D0}.
\item [4:] For all time $t \in [t_0, t_0+ (n-1)T_s]$, compute the safety behavior metric $S_{{R_i}}(t)$ in \eqref{eq10} and $\lambda _i({p_i}(t))$ in \eqref{eq12a}.
\item [5:] Then, evaluate the conditions in \eqref{eq40} and \eqref{eq30}. If both hold true $\forall t \in [t_0, t_0+ (n-1)T_s]$, then agent $i \in \mathcal{V}/\mathcal{V}_{f}$ is identified as adversarial at time $t=t_0+ (n-1)T_s$, i.e., $i \in \mathcal{A}_{s}$.
\item [6:] \textbf{end procedure}
\end{enumerate}
\end{algorithm}

\noindent
of inter-agent safety constraints in \eqref{eq32d} can happen {either due to the behavior of agent $i$ or neighbor agent $j$}. Based on Proposition 2, if $ {{L}_{f_k^p}}{\lambda _k}({p_k}(t)) + {{L}_{g_k^p}}{\lambda _k}({p_k}(t)){u_k}(t) \ge 0, \,\, \forall t \in [t_0, t_0+ (n-1)T_s]$ is satisfied, (which shows that agent $k \in \{i,j\}$ is deviating from its desired goal-reaching behavior) along with the condition  \eqref{eq30}, (which shows that the trajectory of agent $i$ is converging to the boundary of unsafe region at $t=t_0+ nT_s$), then the agent $i$ detects any agent $k$ (i.e., either itself $i$ or any neighbor agent $j$) as adversarial at time $t=t_0+ (n-1)T_s$. This completes the proof.} 
\end{proof}

\noindent

\vspace{-0.3cm}

\begin{remark}
Note that to ensure safety for intact neighbors, one needs to proactively detect agent $i$ as adversarial at $t=t_0+ (n-1)T_s$ if the condition in \eqref{eq30} and $ {{L}_f}{\lambda _i}({p_i}(t)) + {{L}_g}{\lambda _i}({p_i}(t)){u_i}(t) \ge 0$ are satisfied, $\forall t \in [t_0, t_0+ (n-1)T_s]$, and takes proactive action in time window $ [t_0+(n-1)T_s, t_0+ nT_s]$. Otherwise, the adversarial agent $i$ violates safety with its intact neighbor, i.e., $h_{i}^{S} (\vec{p} (t)) > 0,\,\,\forall t > t_0+ nT_s$.
\end{remark}

\vspace{-0.5cm}

\subsection{Task Behavior Metric and Proactive Adversary Detection}

{In this subsection, we first design a metric to capture the behavior of neighboring agents in formation, and then present the result on the detection of an adversarial agent among the set of agents ${{\cal V}_f} \subseteq{\cal V}$.} Similar to $\lambda _i({p_i}(t))$ in \eqref{eq12a}, the goal reaching behavior for a collaborative task among the set of agents ${{\cal V}_f} \subseteq{\cal V}$ in the form of formation  can also be encoded in the following metric 
\vspace{-0.3cm}
\begin{align}\label{m2}
{\lambdabar_f}({\bar p}(t),{\bar p}(0),{G_f}) = \frac{{{{\left\| {{\bar p}(t) - {G_f}} \right\|}^2}}}{{{{\left\| {{\bar p}(0) - {G_f}} \right\|}^2}}} 
=\zeta_f \bar{V}_{f}^{g}(\bar{p}(t)),\,\,\forall t \geq 0,
\vspace{-0.2cm}
\end{align}
where $\zeta_f=\frac{1}{{\left\| {{\bar p}(0) - {G_f}} \right\|}^2}$, $V_{f}^{g}(\bar p(t))$ is defined in \eqref{s6a} and ${\bar p}(0)$ denotes the initial centroid position of agents in formation with $\bar{p}(t) = \frac{1}{{\left| {{{\cal V}_f}} \right|}}\sum\limits_{i \in {{\cal V}_f}} {{p_i}(t)}$. Note that $0\leq  {\lambdabar_f}({\bar p}(t),{\bar p}(0),{G_f}) \leq 1$ under normal behavior of the agents in the formation.

\begin{proposition}
Consider the agent dynamics \eqref{eq1} along with the QP control \eqref{eq4}. For ${\lambdabar_f}({\bar p}(t))$  defined in \eqref{m2},  if it holds that   
\vspace{-0.25cm}
\begin{align}
\begin{gathered}
 (({{L}_{f_i^p}}{\lambdabar_f}({\bar p}(t)) + {{L}_{g_i^p}}{\lambdabar_f}({\bar p}(t)){u_i}(t)) \ge 0) \lor ({\lambdabar_f}({\bar p}(t))) > 1),
\end{gathered} 
\label{eqm3}
\end{align}
for all $t \in [t_k, t_k+ nT_s]$, then at least one of the agent among the set of agents ${{\cal V}_f} \subseteq{\cal V}$  is adversarial.
\end{proposition}

\begin{proof}
The result follows a similar argument as provided in the proof of Proposition 2.
\end{proof}


\vspace{-0.2cm}

\noindent
\textit{Task Behavior Metric:} {Note that the result of Proposition 3 determines that at least one of the agents among the set of agents ${{\cal V}_f}$  might be adversarial. However, in order to accomplish goal reaching for a collective task among the set of intact agents ${{\cal V}_f}/{\cal A}$, one needs to explicitly detect adversarial agents and reject their contribution in the formation. Thus, we design a \textbf{task behavior metric} for monitoring the behavior of neighboring agents in the formation, and present the result on the detection of adversarial agents among the set of agents ${{\cal V}_f}$. The task behavior metric is defined as:}
\vspace{-0.2cm}
\begin{equation}\label{eq19}
    F_{{R_{ij}}}(t) = \exp {( -( \Theta _i(p_i(t),p_j(t)))^{{n_c}}}),\,\,\forall t \geq 0,
\end{equation}
where
\vspace{-0.2cm}
\begin{equation}\label{eq20}
\Theta_i(p_i(t),p_j(t))=  |\left\| {{p_i}(t) - {{p}_j}(t)} \right\|-{c}_{ij}| 
\end{equation}
with desired formation distance $c_{ij}$ between agent $i$ and its neighbor $j$.

 From  $\Theta_i(p_i(t),p_j(t))$ in \eqref{eq20}, in the absence of attack, once desired group of agents reach formation, then $F_{{R_{ij}}}(t)$ will be always close to one during nominal operation and it goes to zero only if agent $i$ or $j$ is adversarial. Based on the result of task behavior metric, we determine confidence value $C_i(t)$ in \eqref{alg1} for each agent $i \in \mathcal{V}_f$ using Algorithm 2. Note that the confidence value $C_i(t)$ in \eqref{alg1} represents the degree of trustworthiness of each agent $i$ about its own information. In particular, if an agent is adversarial, then $C_i(t) \to 0$, otherwise $C_i(t) \to 1$. Now, we define 
\begin{equation}\label{eq21}
    E_i^{{{\cal F}_i}}(p_i(t),p_j(t))= |\left\| {{p_i}(t) - {p_j}(t)} \right\|-c_{ij}|-\Theta_{i}^{w}(p_i(t),p_j(t)) \leq 0,
\end{equation}
where $\Theta_{i}^{w}(p_i(t),p_j(t))>0$ represents the bound on the formation error for agents without considering any adversary.

\begin{remark}
Note that the formation error bound 
${\Theta}_{i}^{w}(p_i(t),p_j(t))$ in \eqref{eq21} can be designed based on exponential convergence for the formation control \cite{sun2016exponential}. In particular, $\bar{\Theta}_{i}^{w}(p_i(t),p_j(t))=k_1e^{-k_2t}( \left\| {{p_i}(0) - {p_j}(0)} \right\|-c_{ij})$ where $k_1=\sqrt{\frac{c_2}{c_1}}$ and $k_2={\frac{\lambda_r\rho}{2c_1}}$ with design constant $c_1,c_2,\rho \in  \mathbb{R_{+}}$ and $\lambda_r$ as minimum singular value of rigidity matrix corresponding to desired formation \cite{sun2016exponential}. Also, for the sake of brevity, one can select an arbitrary large enough scalar value $\Theta_{i}^{w}(p_i(t),p_j(t))$ such that $\Theta_{i}^{w}(p_i(t),p_j(t))>\bar{\Theta}_{i}^{w}(p_i(t),p_j(t))$ for all time $t$.
\end{remark}

 \begin{algorithm}[!ht]
\caption{Determination of confidence value $C_i(t)$ and identification of Adversarial Agents belong to set $\mathcal{A}_{f}$}
\begin{enumerate}
\item [1:] \textbf{procedure} $\forall i \in \mathcal{V}_f$ 
\item [2:] \hspace{0.2cm} \textbf{for} $i=1:|\mathcal{V}_f|$
\item [3:] \hspace{0.5cm} \textbf{initialize} $index=[\,],\,\,\,\mathcal{A}_{f}=[\,]$
\item [4:] \hspace{0.5cm}  \textbf{for} $j=1:\mathcal{N}_i(t)$
\item [5:] \hspace{1.2 cm}
{From \eqref{eq19} and \eqref{eq21a}, evaluate $F_{{R_{ij}}}(t)$ and  $ F_{{R_i}}^w(t)$}.
\item [6:] \hspace{1.2 cm} \textbf{if} $F_{{R_{ij}}}(t)> F_{{R_i}}^w(t)$ \item [7:]\hspace{2 cm}$index=[index \,\,\, j];$
\item [8:] \hspace{1.2 cm} \textbf{end if} 
\item [9:] \hspace{0.5cm}  \textbf{end for} 
\item [10:] \hspace{0.5cm} Using $index$ from step 7, evaluate confidence
\begin{equation}\label{alg1}
    C_{i}(t) = \exp {( - (\frac{2|index|}{\mathcal{N}_i(t)})^{n_c}}),\,\,\forall t \geq 0, \,\,n_c \in \mathbb{Z}_{>1}.
\end{equation}
\item [11:] \hspace{0.5cm} \textbf{if} $(2|index|)>\mathcal{N}_i(t)$ 
\item[12:] \hspace{1.2 cm} $\mathcal{A}_{f}=[\mathcal{A}_{f}\,\,\,\,\,i];$
\item [13:] \hspace{0.5cm} \textbf{end if}
\item [14:] \hspace{0.2cm}  \textbf{end for}
\item [15:] \textbf{end procedure}
\end{enumerate}
\end{algorithm}

We also define the worst-case collective behavior as
\vspace{-0.1cm}
\begin{equation}\label{eq21a}
    F_{{R_{ij}}}^w(t) = \exp {( - (\Theta_{i}^{w}(p_i(t),p_j(t)))^{{n_c}}}),\,\,\forall t \geq 0.
\end{equation}

\vspace{-0.1cm}
Then, define the error between the worst-case and nominal formation behavior metrics as
\vspace{-0.1cm}
\begin{equation}\label{c2}
\gamma _i^F(t) = \left\| {F_{{R_{ij}}}^{w}(t) - F_{{R_i}}^{n}(t)} \right\|,
\end{equation}
with ${F_{{R_i}}^n(t)} = 1$ as the nominal task behavior metric. From $\Theta_i(p_i(t),p_j(t))$ in \eqref{eq20}, in the absence of any adversary, once desired group of agents reach formation, then $F_{{R_i}}(t)=1$ and that shows the nominal behavior of agents in the formation.

 Based on the design of $\Theta_{i}^{w}(p_i(t),p_j(t))$, $E_i^{{{\cal F}_i}}(p_i(t),p_j(t)) \leq 0$ always holds true for each $i \in \mathcal{V}_f$ if $\mathcal{A} \cap \mathcal{V}_f =\emptyset $. Thus, in the following result, based on the designed task behavior metric, one can detect the adversarial agents among the set of collaborative agents ${\cal V}_f$.
 \vspace{-0.2cm}
 \begin{theorem}
  Consider the agent dynamics \eqref{eq1} along with the QP based control  \eqref{eq4}. For the task behavior metric $F_{{R_{ij}}}(t)$ in \eqref{eq19} and  $\gamma _i^F (t)$ in \eqref{c2}, if an agent $i \in \mathcal{V}_f$ satisfies   
\vspace{-0.2cm}
\begin{equation}
{\left\| {{F_{{R_{ij}}}(t)} - F_{{R_i}}^{n}(t)} \right\|} > {\gamma _i^F},
\label{eq22}
\end{equation}
for more than $\frac{\mathcal{N}_i(t)}{2}$ neighbors at some time $t \geq 0$, then the  agent $i$ is detected as adversarial among the set of agents $\mathcal{V}_f$.
 \end{theorem}

\begin{proof}
Based on the statement of theorem, if \eqref{eq22} holds, then from \eqref{c2}, with ${F_{{R_i}}^n(t)} = 1$ as the nominal task behavior metric, $F_{{R_{ij}}}(t)> F_{{R_i}}^w(t)$ also holds for more than $\frac{\mathcal{N}_i(t)}{2}$ neighbors at some time $t \geq 0$. Based on task behavior metrics $F_{{R_{ij}}}(t)$ and $F_{{R_i}}^w(t)$ in \eqref{eq19} and \eqref{eq21a}, one has formation error in  \eqref{eq21} as
\begin{equation}\label{eqth5}
  E_i^{{{\cal F}_i}}(p_i(t),p_j(t))= |\left\| {{p_i}(t) - {p_j}(t)} \right\|-c_{ij}|-\Theta_{i}^{w}(p_i(t),p_j(t)) > 0,
\end{equation}
for more than $\frac{\mathcal{N}_i(t)}{2}$ neighbors. That means the agent violates the formation error bound with more than half of its neighbors and this is only possible if agent $i$ itself is adversarial because we assumed at max half of neighbors can be adversarial. Thus, the agent $i$ is detected as an adversarial agent. 
\end{proof}


\vspace{-0.25cm}

 Note also that the set of adversarial agents $\mathcal{A}=\mathcal{A}_s \cup \mathcal{A}_f$ can be determined based on Algorithm 1 and Algorithm 2. The set of adversarial agents $\mathcal{A}_s$ is computed based on the results presented for safety behavior metric in Theorem 4. Similarly, the set of adversarial agents $\mathcal{A}_f$ is evaluated based on the task behavior metric in Theorem 5. After identifying the set of adversarial agents $\mathcal{A}$, we leverage the adversary detection results along with presented behavior metrics for the design of resilient QP in the next section.

\section{Resilient Controller Design}




 This section presents the formulation of resilient quadratic program (QP) to compute a control input $u_{i}(t) $ for each agent $i$ to solve Problem 1 in the presence of the set of adversarial agent $\mathcal{A}$.  Let $\vec z_r= [z_1^T,z_2^T,\dots,z_{N-|\cal{A}|}^T]^T$ be a column vector with   $z_{i} =[u_{i} ,{\delta _{{i_1}}},{\delta _{{i_2}}},{\delta _{{i_3}}},{\delta _{{i_4}}},\{ {\delta _{ij}}\} ]^{T} \in \mathbb{R}^{m_{i} +4+{N}_{i} } $ with ${N}_{i} =\left|{\rm {\mathcal N}}_{i} \right|,\,\,\,\, \forall i\in \cal{V}/\cal{A}$ as its elements. Consider the following optimization problem 
\vspace{-0.2cm}
\begin{subequations}
\begin{align}
\begin{split}\label{QP2a}
 \mathop {\min \,}\limits_{{u_i},{\delta _{{i_1}}},{\delta _{{i_2}}},{\delta _{{i_3}}},\{ {\delta _{ij}}\} ,\,\,i \in {\cal V}/\mathcal{A}} \,\,\,{{\vec z_r}^T}H\vec z_r + F\vec z_r\\
\end{split}\\
\begin{split}\label{QP2b}
s.t.\,\,\,\,\,\,{A_i}{u_i} \le {b_i}\\
\end{split}\\
\begin{split}\label{QP2c}
{L_{{f_i^p}}}{V}_i^g + {L_{{g_i^p}}}{V}_i^g{u_i} \le  - {\delta _{{i_1}}}{V}_i^g,\,\,\forall i \in \mathcal{V}/\mathcal{V}_f\\
\end{split}\\
\begin{split}\label{QP2c1}
{{L}_{f_i^p}}{\bar{V}_f^r} + {{L}_{g_i^p}}{\bar{V}_f^r}{u_i}(t) \le - {\delta _{{i_2}}}\bar{V}_f^r\\
\end{split}\\
\begin{split}\label{QP2d}
{L_{{f_i^p}}}{h}_i^{o_j} + {L_{{g_i^p}}}{h}_i^{o_j}{u_i} \le  - {\delta _{{i_3}}}{h}_i^{o_j},\,\,\,\,\forall o_j \in {{\cal O}_i}\\
\end{split}\\
\begin{split}\label{QP2e}
{L_{{f_i^p}}}\bar{h}_i^{{{\cal F}_i}} + {L_{{g_i^p}}}\bar{h}_i^{{{\cal F}_i}}{u_i} \le  - {\delta _{{i_4}}}\bar{h}_i^{{{\cal F}_i}},\forall i \in {{\cal V}_f}\\
\end{split}\\
\begin{split}\label{QP2f}
{L_{{f_i^p}}}\bar{h}_{ij}^S + {L_{{g_i^p}}}\bar{h}_{ij}^S{u_i} + \pi (\bar h_{ij}^S)  \le  - {\delta _{ij}}\bar{h}_{ij}^S,\,\,\forall j \in {{\cal N}_i},
\end{split}
\end{align}
\label{QP2}
\end{subequations}

\vspace{-0.4cm}
\noindent
where $H=diag\{H_i\}$ with $H_{i} =diag\{ \{ w_{u_{l} }^{i} \} ,w_{1}^{i} ,w_{2}^{i},w_{3}^{i},w_{4}^{i},\{ w_{n_{p} }^{i} \} \}$ denotes  a diagonal matrix with positive weights $w_{u_{l} }^{i} ,w_{1}^{i} ,w_{2}^{i},w_{3}^{i},w_{4}^{i}, w_{n_{p} }^{i} >0$ for each $p\in {\rm {\mathcal N}}_{i} ,$  and similarly, $F=diag\{F_i\}$  with $\,\,F_{i} =[0_{m_{i} }^{T} \, \, \,q^{i} \,\, \, 0_{\mathcal{N}_i +3}\, ]$ where $q^{i} >0$ and $0_{k} \in \mathbb{R}^{k}$ denotes a column vector consisting of zeros. {Moreover, based on the task behavior metric $F_{{R_{ij}}}(t)$ in \eqref{eq19}, the function $ \bar{h}_i^{{{\cal F}_i}}(t)$ in \eqref{QP2e} is defined in the resilient form as
\vspace{-0.2cm}
\begin{equation} \setcounter{equation}{75} 
  \begin{array}{l}
\bar h_i^{{F_i}}(t) = \left\| {{p_i}(t) - \hat p_i^*(t)} \right\|,\\
\end{array} \label{eq74}
\end{equation}
with  $\hat{p}_i^*(t) = \frac{1}{{\left| {{{\cal N}^{ F}_i}(t)} \right|}}\sum\limits_{j \in {{\cal N}_i}} {  F_{{R_{ij}}}(t)({p_j}(t) + {c_{ji}})}$ and $\left| {{{\cal N}^{ F}_i}(t)} \right|=\sum\limits_{j \in {{\cal N}_i}} F_{{R_{ij}}}(t)$. Similarly, the function $\bar h_{ij}^S$ in \eqref{QP2f} is defined as
\vspace{-0.3cm}
\begin{equation}
  \begin{array}{l}
\bar h_{ij}^S = h_{ij}^S + \frac{{\eta ({p_{ij}})}}{n},
\end{array} \label{eq74a}
\end{equation}
and
\vspace{-0.35cm}
\begin{equation}\label{eq74b}
\pi (\bar h_{ij}^S(t)) = \left\{ \begin{array}{l}
{{L}_{{f_j^p}}}\bar h_{ij}^S + {{L}_{{g_j^p}}}\bar h_{ij}^Su_j^{}(t),\,\,if\,\,j \notin  \mathcal{A},\\
{{L}_{{f_j^p}}}\bar h_{ij}^S + {{L}_{{g_j^p}}}\bar h_{ij}^Su_j^{max}(t),\,\,if\,\,j \in  \mathcal{A}.
\end{array} \right.
\end{equation}
where $u_j^{max}(t)$ worst case adversarial control action defined in \eqref{eq8c}. Also, based on the confidence value $C_i(t)$ in \eqref{alg1}, the resilient CLF for collaborative goal reaching is defined as
\vspace{-0.1cm}
\begin{equation}\label{eq73}
\bar{V}_f^r(t)= \left\| \bar{p}^r(t) - G_f \right\|^2 \to 0
\end{equation}
as $t \to \infty$ with $G_f$ as goal point or goal region, $\bar{p}^r(t) = \frac{1}{{\cal N}^{r}(t)} \sum\limits_{i \in {{\cal V}_f}} {C_{i}(t){p_i}(t)}$ and ${\cal N}^{r}(t)= \sum\limits_{i \in {{\cal V}_f}} C_{{i}}(t) $  for collaborative goal reaching with the set of agents ${{\cal V}_f} \subseteq{\cal V}$ in the form of formation.}

Now, in the following theorem, we present the result that solves Problem 1 with objectives A.1-A.4 for each agent $i \in \mathcal{V}/\mathcal{A}$. Let the solution of \eqref{QP2} be represented by  $\vec z_r^*= [z_1^{{*}^{T}},z_2^{{*}^{T}},\dots,z_{N-|\cal{A}|}^{{*}^{T}}]^T$ where $z_{i}^{*}(.) =[u_{i}^{*}(.) ,{\delta _{{i_1}}^{*}(.)},{\delta _{{i_2}}^{*}(.)},{\delta _{{i_3}}^{*}(.)},{\delta _{{i_4}}^{*}(.)},\{ {\delta _{ij}^{*}(.)}\} ]^{T}$.

\begin{theorem}
Consider the agent dynamics \eqref{eq1}. Then, 
\begin{enumerate}
\item the resilient QP in \eqref{QP2} is feasible for each intact agent  $i \in \mathcal{V}/\mathcal{A}$, and

\item the resilient QP in \eqref{QP2} solves Problem 1 for each intact agent $i \in \mathcal{V}/\mathcal{A}$.
\end{enumerate}
\end{theorem}

\begin{proof}

Since, we have $\bar{V}_f^r>0$ in \eqref{eq73} for all $\bar{p}^r_i(t) \notin G_f$ and $t \geq 0$. One can select $u_i=u_{i}^{*} \in {\rm {\mathcal U_i}}$ and define 
\begin{equation}\label{73}
\delta _{{i_2}}=-\frac{{{L}_{f_{i}}}{\bar{V}_f^r} + {{L}_{g_{i}}}{\bar{V}_f^r}{u_{i}^{*}}(t)}{\bar{V}_f^r},
\end{equation}
and it can be explicitly defined for all $\bar{p}^r_i(t) \notin G_f$, such that \eqref{QP2c1} satisfies the equality condition. We know that $\bar{h}_{ij}^S<0$ for all $p_i(t) \in int(\bar S_{i}^{s})$ and for all $t \geq 0$, where the set $\bar S_{i}^{s}$ is defined in \eqref{s2}. Then, one can chose $u_i=u_{i}^{*} \in {\rm {\mathcal U_i}}$ and define 
\begin{equation}
\delta _{{ij}}=-\frac{{L_{{f_i^p}}}\bar{h}_{ij}^S + {L_{{g_i^p}}}\bar{h}_{ij}^S{u_i} + \pi (\bar h_{ij}^S) }{\bar{h}_{ij}^S},
\end{equation}
and it can be explicitly defined for all $p_i(t) \in int(\bar S_{i}^{s})$, such that \eqref{QP2f} satisfies the equality condition.
Similarly, we know that one can define slack parameters ${\delta _{{i_1}}^{*}},{\delta _{{i_3}}^{*}},{\delta _{{i_4}}^{*}}$ such that \eqref{QP2b} and \eqref{QP2c1}-\eqref{QP2e} are satisfied with equality condition. Therefore, there exists $z_{i}^{*}(.) =[u_{i}^{*}(.) ,{\delta _{{i_1}}^{*}(.)},{\delta _{{i_2}}^{*}(.)},{\delta _{{i_3}}^{*}(.)},{\delta _{{i_4}}^{*}(.)},\{ {\delta _{ij}^{*}(.)}\} ]^{T}$ all constraints in \eqref{QP2} are satisfied and the resilient QP in \eqref{QP2} is feasible for each intact agent $i \in \mathcal{V}/\mathcal{A}$.

Now based on result of part 1, we present the proof of part 2. Since the resilient QP in \eqref{QP2} is feasible for all $i \in \mathcal{V}/\mathcal{A}$, thus there exist $z_{i}^{*}(.) =[u_{i}^{*}(.) ,{\delta _{{i_1}}^{*}(.)},{\delta _{{i_2}}^{*}(.)},{\delta _{{i_3}}^{*}(.)},{\delta _{{i_4}}^{*}(.)},\{ {\delta _{ij}^{*}(.)}\} ]^{T}$ which ensures \eqref{QP2b}-\eqref{QP2f} for all $t>0$. Based on the confidence value $C_{i}(t)$ in \eqref{alg1}, the resilient CLF $\bar{V}_f^r$ for collaborative goal reaching in \eqref{eq73} discards the adversarial agent contribution with centroid with $\bar{p}^r(t) = \frac{1}{{\cal N}^{r}(t)} \sum\limits_{i \in {{\cal V}_f}} {C_{i}(t){p_i}(t)}$ and ${\cal N}^{r}(t)= \sum\limits_{i \in {{\cal V}_f}}C_{i}(t)$ as  based on Algorithm 2 $C_{i}(t) \to 0$  for the adversarial agents. Since there exist $z_{i}^{*}(.) =[u_{i}^{*}(.) ,{\delta _{{i_1}}^{*}(.)},{\delta _{{i_2}}^{*}(.)},{\delta _{{i_3}}^{*}(.)},{\delta _{{i_4}}^{*}(.)},\{ {\delta _{ij}^{*}(.)}\} ]^{T}$  for all $i \in \mathcal{V}/\mathcal{A}$ which satisfies constraint  \eqref{QP2c1}, this means the centroid of intact agents among the set of agents $\mathcal{V}_f$ exponentially reaches the goal point. With similar argument, the constraints in \eqref{QP2b}-\eqref{QP2c} and \eqref{QP2d}-\eqref{QP2f} are also satisfied and thus the resilient QP in \eqref{QP2} solves Problem 1 for each intact agent $i \in \mathcal{V}/\mathcal{A}$. This completes the proof.
\end{proof}

\noindent


\begin{remark}
{Note that based on the result in Theorem 6,  the feasibility of the designed resilient QP in \eqref{QP2} depends on the optimization parameters ${\delta _{{i_1}}^{*}(.)},{\delta _{{i_2}}^{*}(.)},{\delta _{{i_3}}^{*}(.)},{\delta _{{i_4}}^{*}(.)}, \delta _{ij}^{*}(.) $. These optimization parameters are similar to feasibility parameters presented in \cite{zeng2021safety,powell2016towards,xiao2020sufficient} to solve the conflicts among constraints and to ensure the feasibility of the CBF-CLF based QP's. Interested readers can refer to \cite{zeng2021safety, xiao2020sufficient} for more details on feasibility analysis.}
\end{remark}

\vspace{-0.2cm}

\section{Numerical Case Studies}
In this section, we present two case studies to demonstrate the efficacy of the presented theoretical contributions. In the first case, we consider a multi-agent problem with objectives to visit some regions (goal reaching) while maintaining safety constraints (i.e., inter-agent and agent-to-obstacle safety), despite the presence of an adversarial agent. In the second case, we consider the multi-agent formation problem under some desired specifications, where the aim is to maintain formation among the set of collaborative agents and visit some regions while maintaining safety constraints with collaborative goal reaching even in the presence of adversarial agents.

\subsubsection{Case 1}
\begin{figure}[!t] 
\begin{subfigure}{0.23\textwidth}
\centering{\includegraphics[width=1\columnwidth] {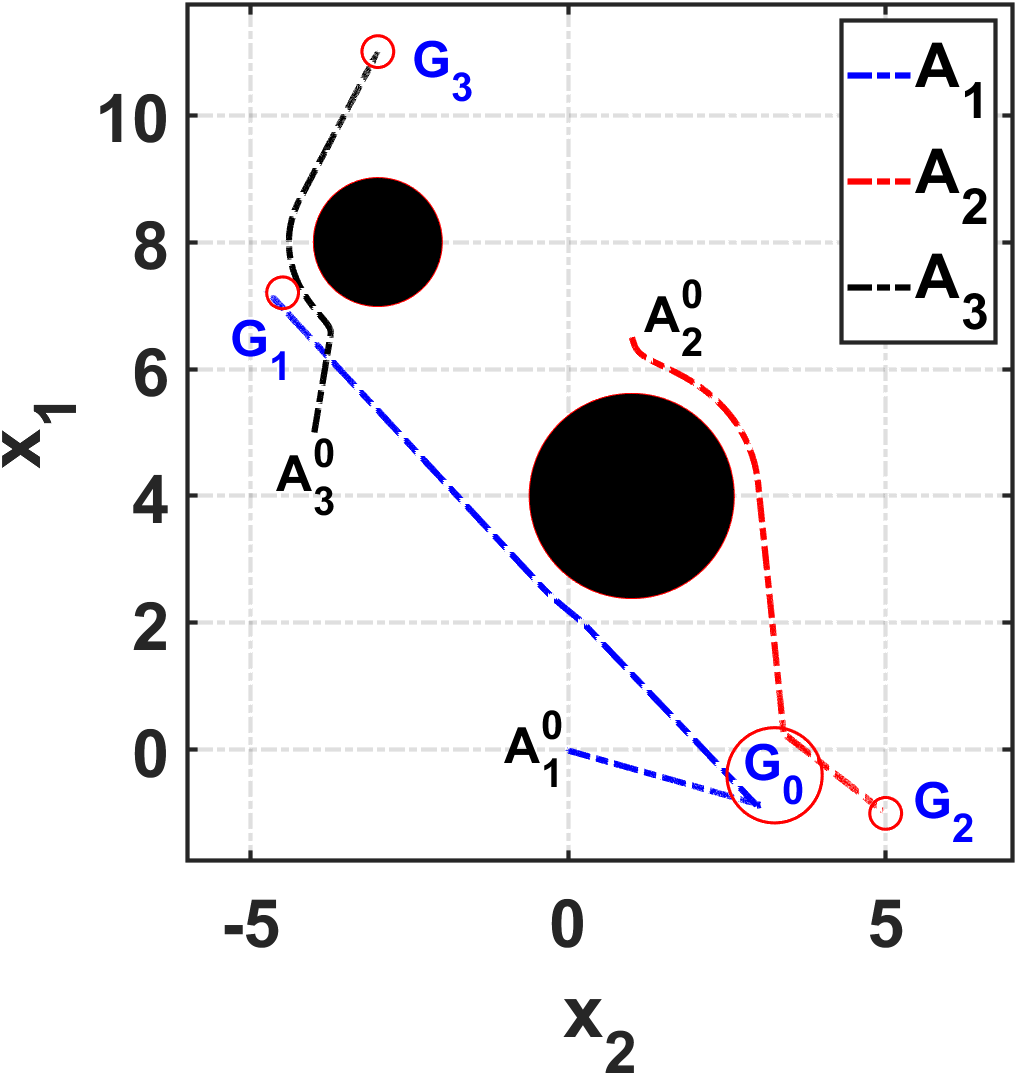}}
\caption{}
\label{fig2a}
\vspace{-0.2cm}
\end{subfigure}
\begin{subfigure}{0.23\textwidth}
\centering{\includegraphics[width=1\columnwidth] {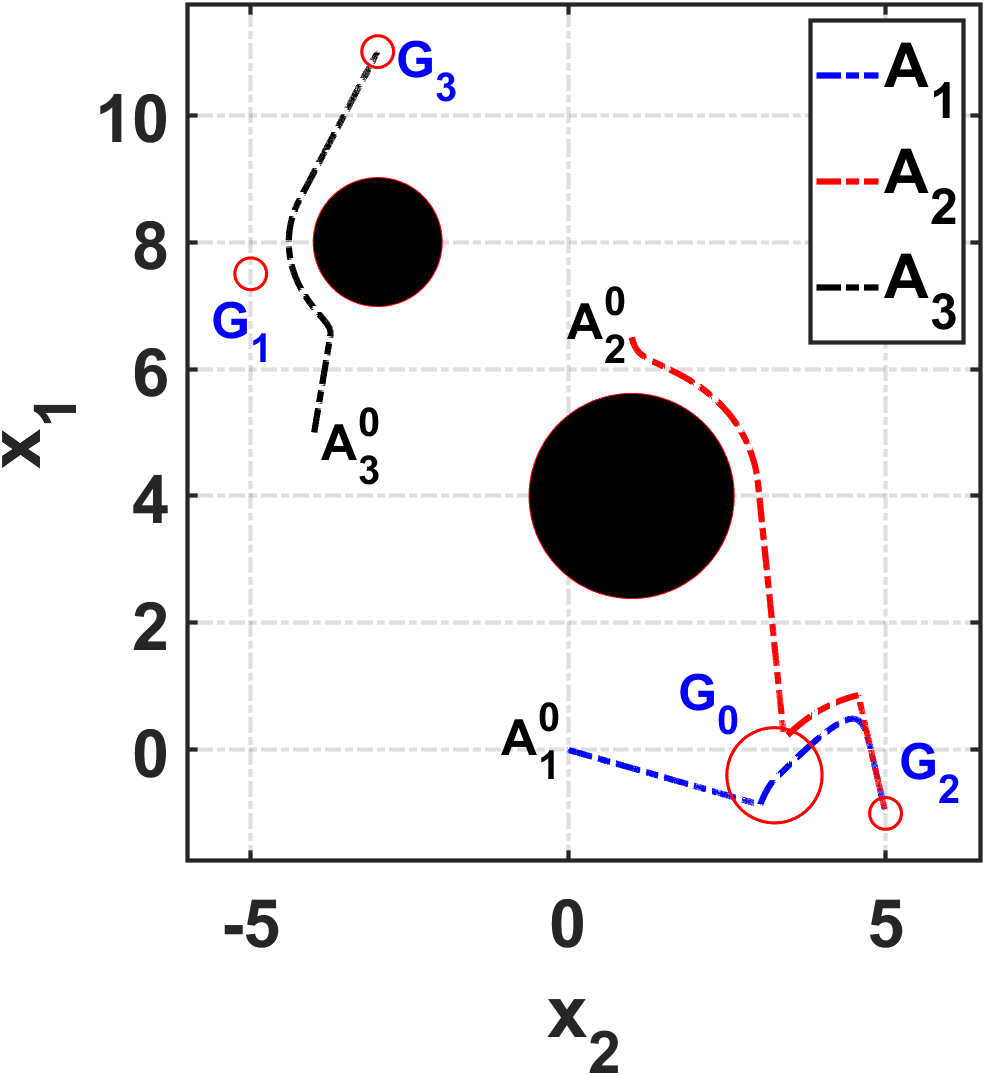}}
\caption{}
\label{fig2b}
\vspace{-0.2cm}
\end{subfigure} 
\caption{Agent behaviors under the desired specification: (a) without adversaries. (b) when agent 1 performs adversarial chasing towards agent 2 for $t>800s$. }
\label{fig2}
\vspace{-0.3cm}
\end{figure}

\begin{figure}[!t] 
\begin{subfigure}{0.24\textwidth}
\centering{\includegraphics[width=1.13\columnwidth] {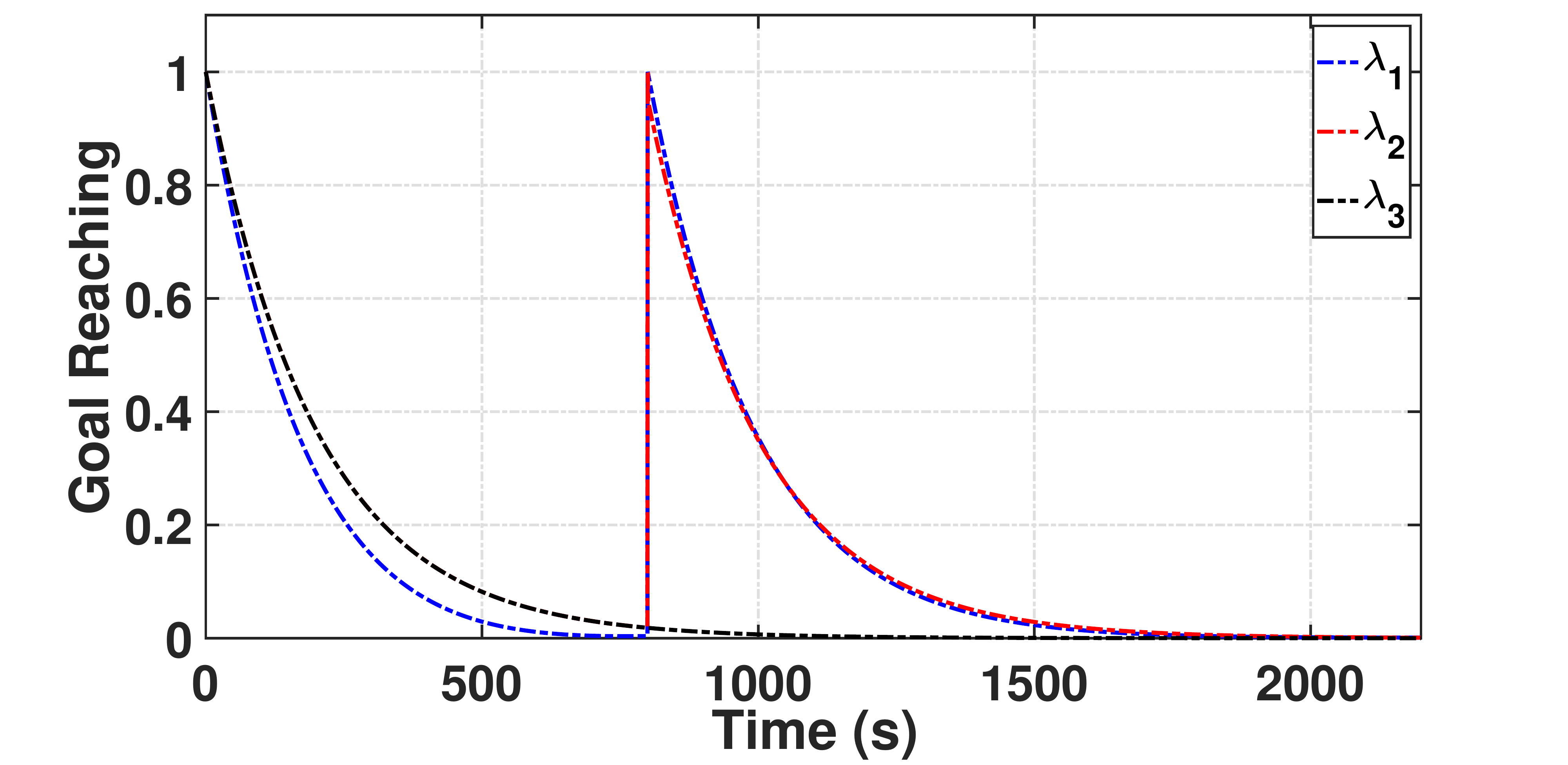}}
\caption{}
\label{fig3a}
\vspace{-0.2cm}
\end{subfigure}
\begin{subfigure}{0.24\textwidth}
\centering{\includegraphics[width=1.13\columnwidth] {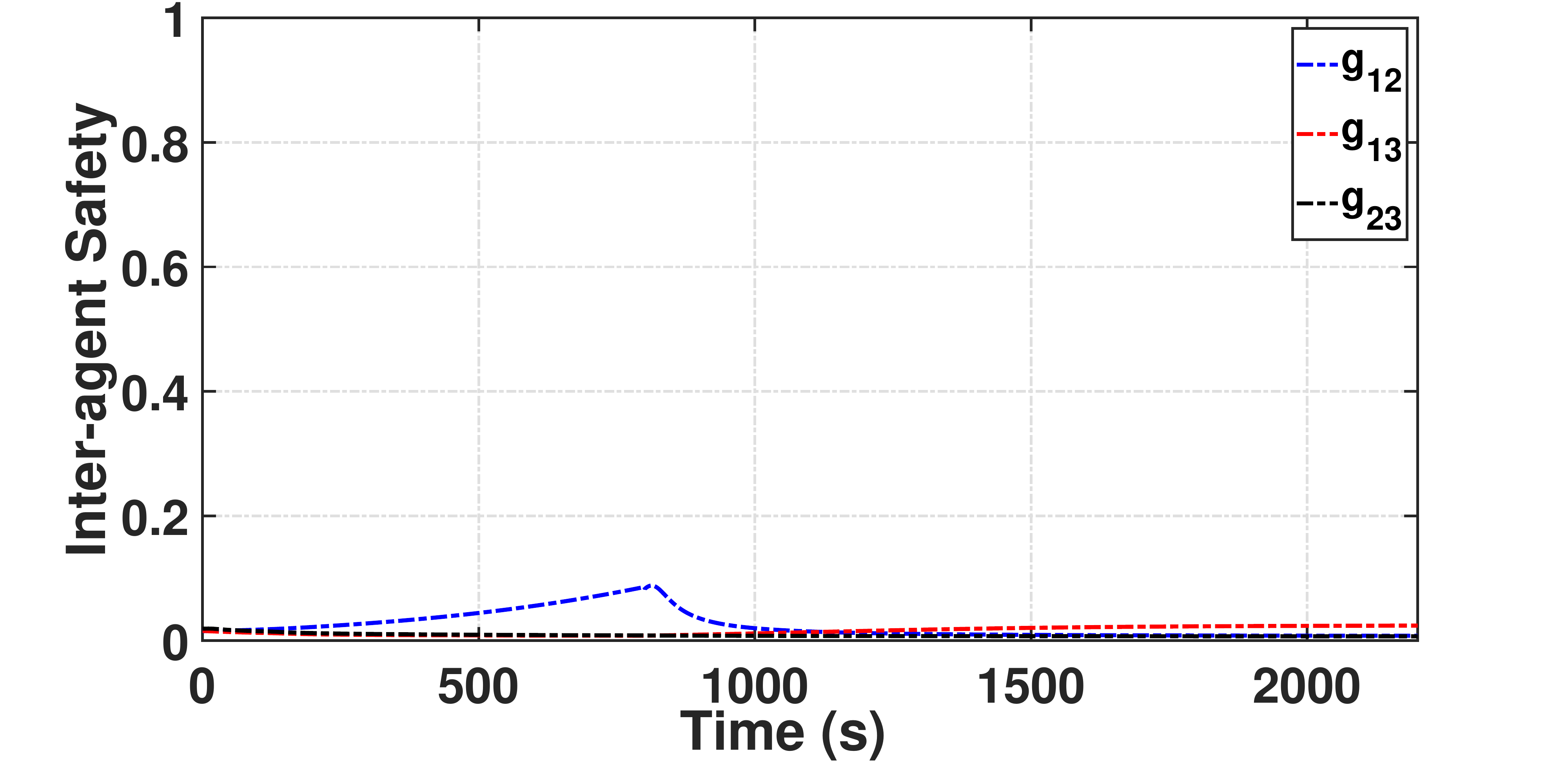}}
\caption{}
\label{fig3b}
\vspace{-0.2cm}
\end{subfigure} 
\caption{(a) Normalized CLF for goal-reaching behavior in the absence of adversaries. (b) Inter-agent safety behavior in the absence of adversaries.}
\label{fig3}
\vspace{-0.3cm}
\end{figure}

We consider a network of three agents with the following linearized unicycle dynamics
\vspace{-0.2cm}
\[\left[ {\begin{array}{*{20}{c}}
{\dot y_1^i}\\
{\dot y_2^i}\\
{{{\dot \theta }^i}}
\end{array}} \right] = \left[ {\begin{array}{*{20}{c}}
1&0\\
0&1\\
{\frac{{ - \sin ({\theta ^i})}}{b}}&{\frac{{\cos ({\theta ^i})}}{b}}
\end{array}} \right]\,\left[ {\begin{array}{*{20}{c}}
{u_1^i}\\
{u_2^i}
\end{array}} \right],\,\,\,\,\forall i \in \{ 1,2,3\} \]
where $y_1^i = {x^i} + b\cos ({\theta ^i})$ and $
y_2^i = {y^i} + b\sin ({\theta ^i})$ with ${\left[ {\begin{array}{*{20}{c}}
{{x^i}}&{{y^i}}
\end{array}} \right]^T}$ and $\theta ^i$ as the position vector and orientation of agent $i$. For the linearized unicycle model, the control input transformation is given by 
\vspace{-0.2cm}
\[\left[ {\begin{array}{*{20}{c}}
{v_{}^i}\\
{w_{}^i}
\end{array}} \right] = {\left[ {\begin{array}{*{20}{c}}
{\cos ({\theta ^i})}&{ - b\sin ({\theta ^i})}\\
{\sin ({\theta ^i})}&{b\cos ({\theta ^i})}
\end{array}} \right]^{ - 1}}\left[ {\begin{array}{*{20}{c}}
{u_1^i}\\
{u_2^i}
\end{array}} \right],\,\,b > 0\]
Under normal operation, the multi-agent has following  desired specifications or objectives $\phi  = {\diamondsuit _{[0,800]}}({\Upsilon _1}^{{G_0}} \wedge {\Upsilon _2}^{{G_0}}\,) \wedge {\diamondsuit _{[800,2200]}}({\Upsilon _1}^{{G_1}} \wedge {\Upsilon _2}^{{G_2}}) \wedge {\diamondsuit _{[0,2200]}}({\Upsilon _3}^{{G_3}}) \wedge \square_{[0,2200]}{\phi _s} $ with ${\Upsilon _i}^{{G_r}} = \left\| {{p_i} - {G_r}} \right\| < \delta_r,\,\,\,\forall r \in \{ 0,1,2,3\}$ and ${\phi _s} = ({\left\| {{p_i} - {p_j}} \right\|^{}} > 0.1) \wedge ({\left\| {{p_i} - {c_{{o_j}}}} \right\|^{}} > 0.4),\, \forall o_j \in \{o_1,o_2\}$ as desired goal reaching and safety specification for agent $i$, respectively. $\delta_r=0.25,\,\,\,\forall r \in \{1,2,3\}$ and $\delta_0=0.75$ for goal regions. 
In particular, the objective for Agents 1 and 2 is to eventually reach the goal location $G_0$ and perform some task between time duration $t \in [0, 800]$, then Agent 1 and Agent 2 are supposed to reach their desired goal locations $G_1$ and $G_2$ over time duration $t \in (800, 2200]$ while maintaining inter-agent and agent-obstacle constraints. Similarly, Agent 3 has to reach its desired goal location $G_3$ over time duration $t \in [0, 2200]$. Figure \ref{fig2a} shows the normal agent's behavior for the desired specification $\phi$. Based on the normalized CLF in \eqref{eq12a} and inter-agent safety metric in \eqref{eq11a}, the goal-reaching and inter-agent safety behavior of agents are shown in Figures \ref{fig3a} and \ref{fig3b}, respectively. One can see that in absence of adversarial agents, all intact agents follow the desired behavior under the control action obtained from the nominal QP \eqref{eq4}. {In Figure \ref{fig2a}, $A_1^0$, $A_2^0$ and $A_3^0$ denote the initial positions of respective agents.}

 \begin{figure}[!t]
\centering{\includegraphics[width=0.96\columnwidth] {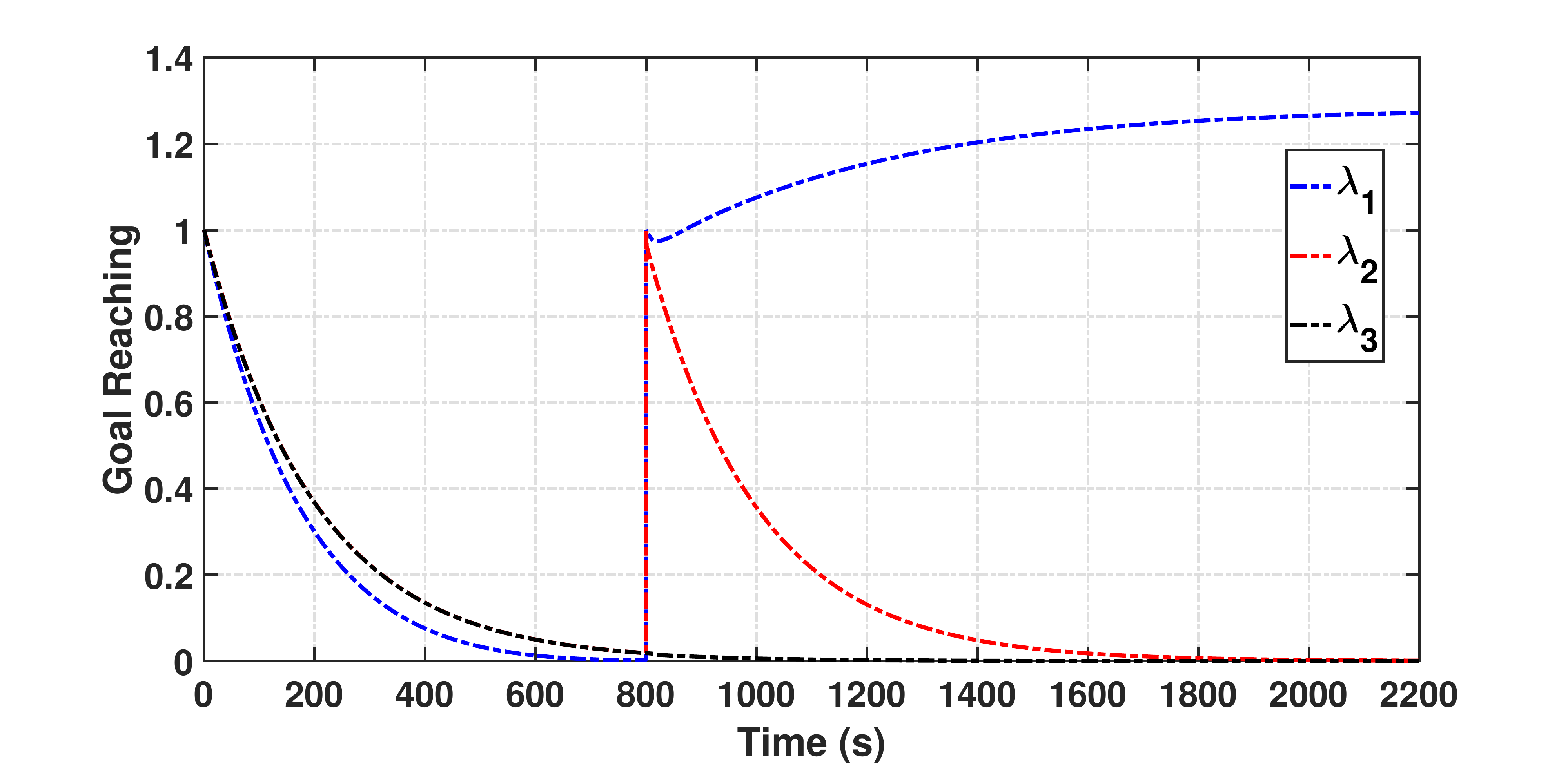}}
 \caption{Normalized CLF for goal-reaching behavior under adversarial chasing.}
\label{fig6}
\vspace{-0.45cm}
\end{figure}

 \begin{figure}[!t]
\centering{\includegraphics[width=0.96\columnwidth] {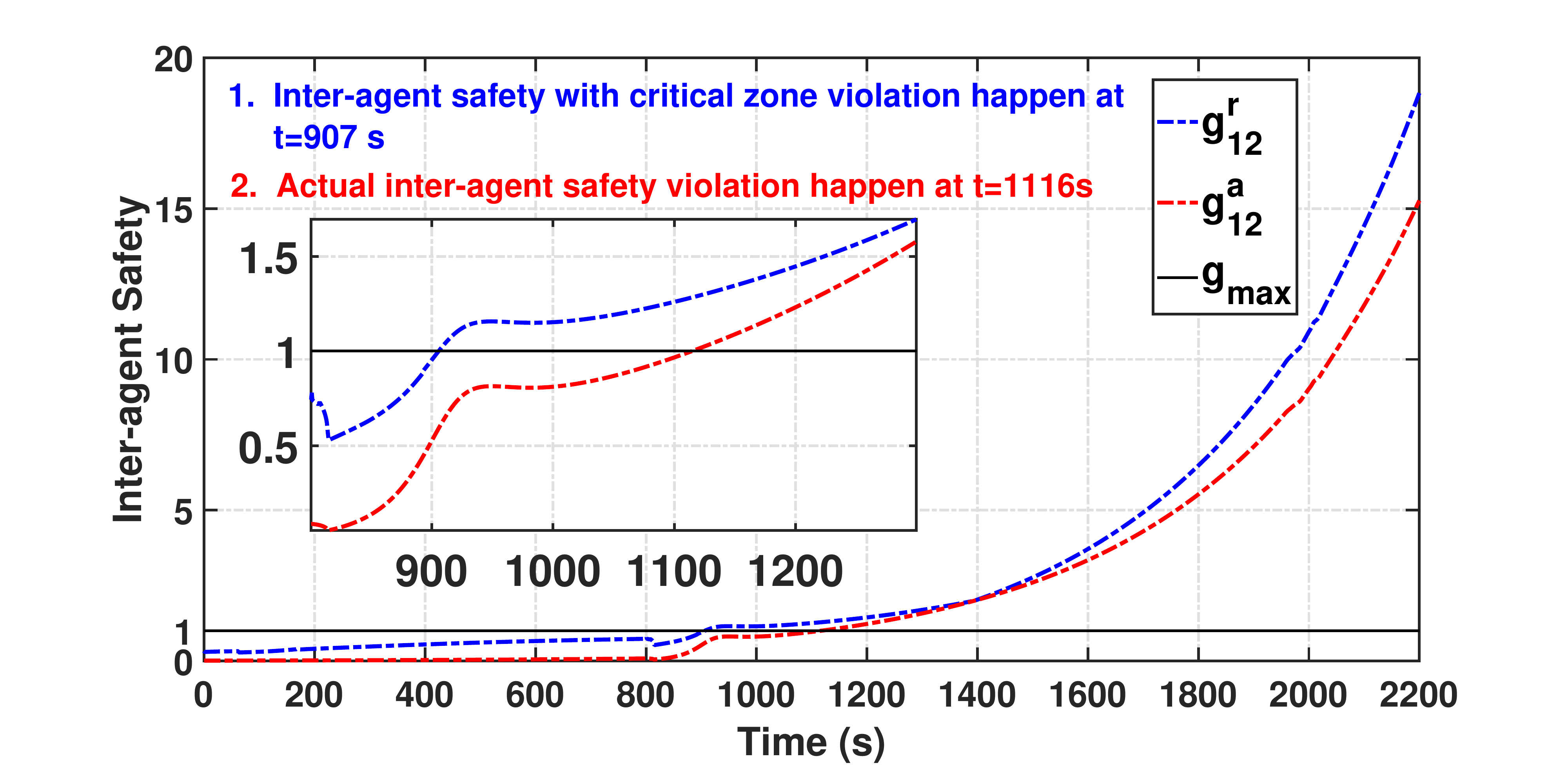}}
 \caption{Inter-agent safety behavior under adversarial chasing.}
\label{fig7}
\vspace{-0.45cm}
\end{figure}

 \begin{figure}[!t]
\centering{\includegraphics[width=0.96\columnwidth]{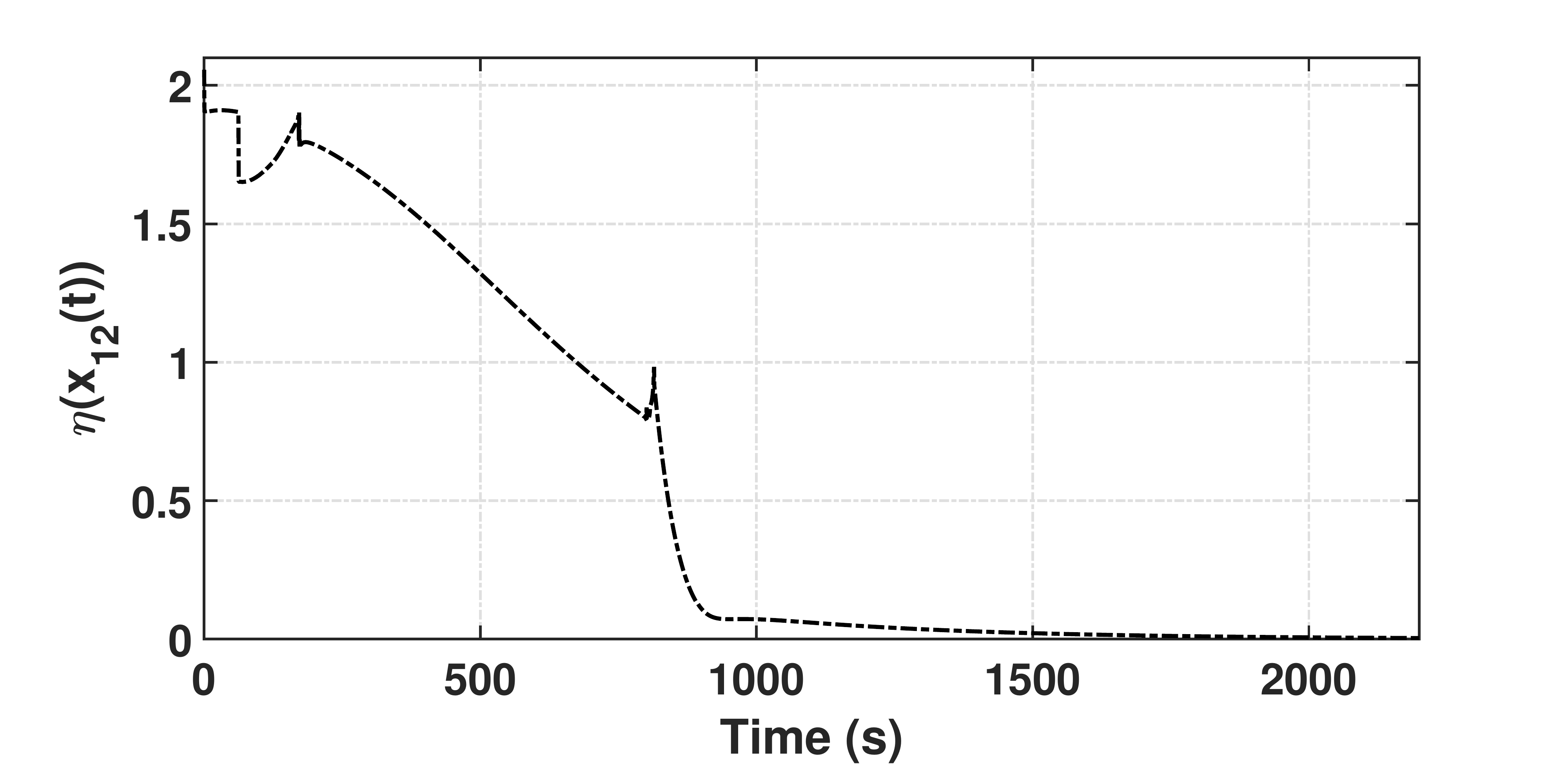}}
 \caption{Critical zone between Agents 1 and 2.}
\label{fig8}
\vspace{-0.25cm}
\end{figure}


\begin{figure}[!t]
\centering{\includegraphics[width=0.65\columnwidth] {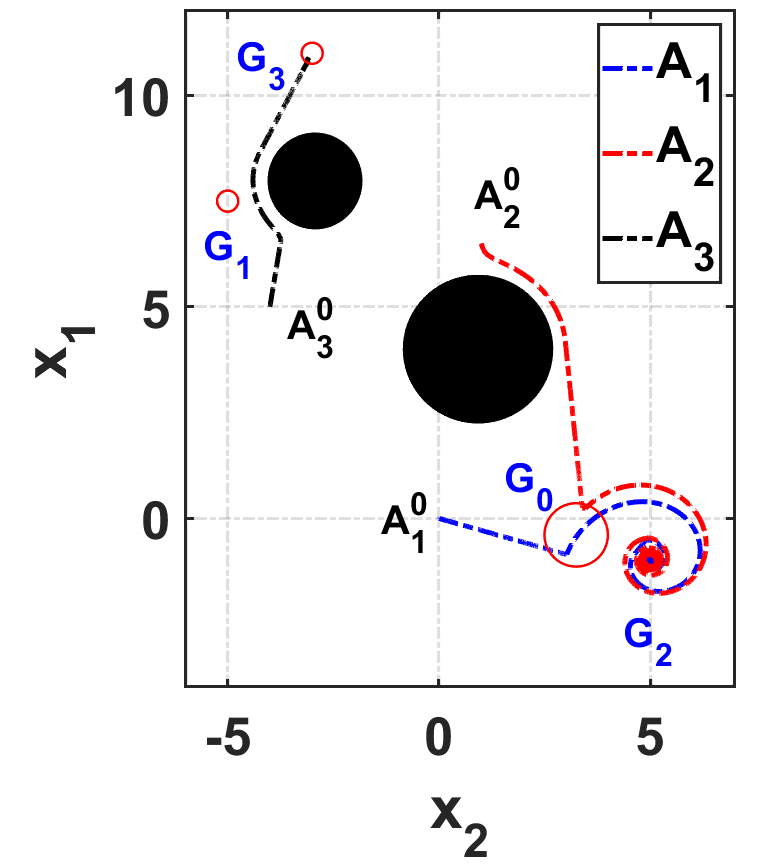}}
\caption{Agent behavior under the resilient QP in \eqref{QP2}: agent 1 chases agent 2 for $t>800s$.}
\label{fig10}
\vspace{-0.4cm}
\end{figure}

 \begin{figure}[!t]
\centering{\includegraphics[width=0.91\columnwidth] {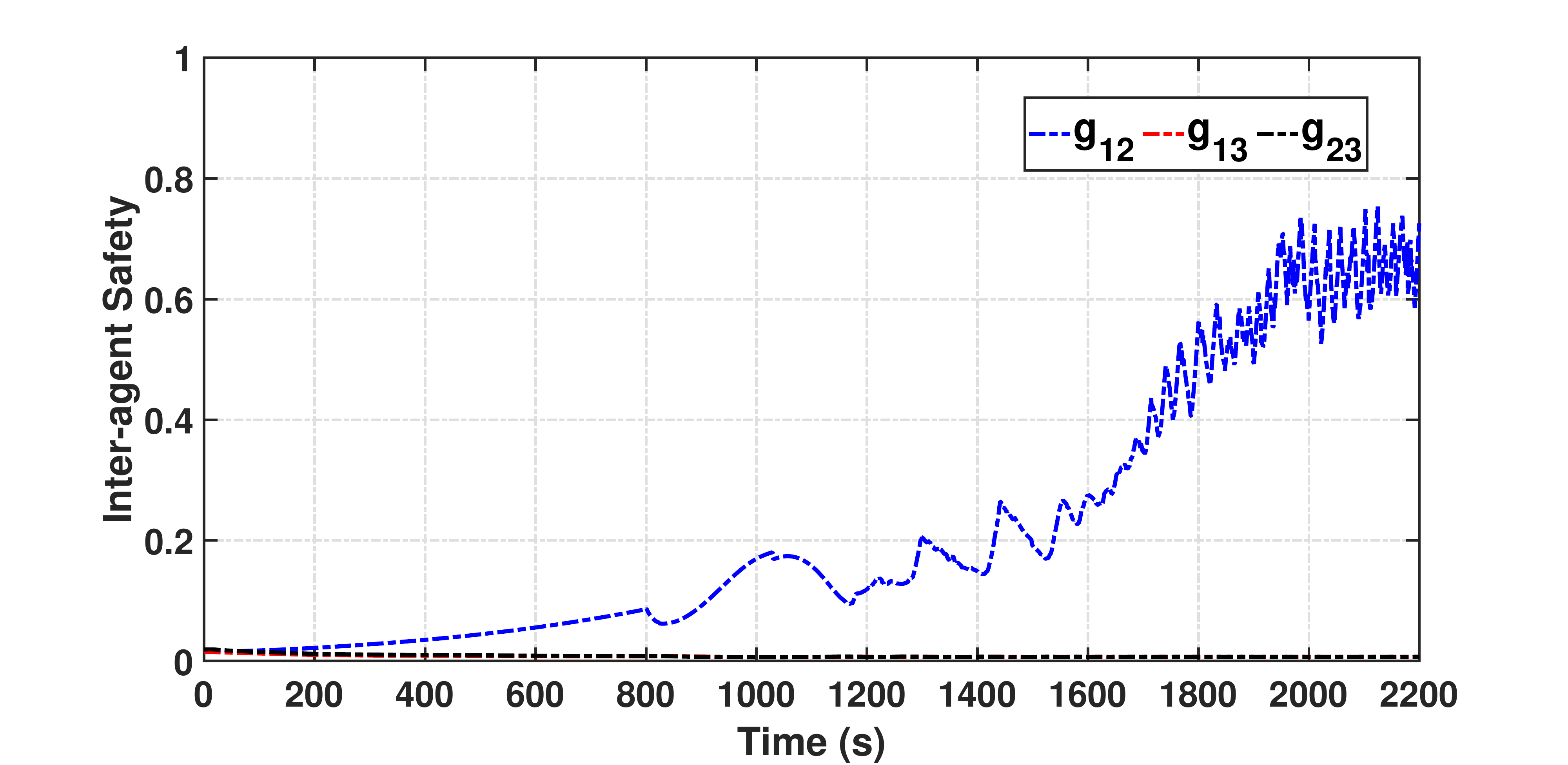}}
 \caption{Inter-agent safety behavior under the resilient QP in \eqref{QP2}.}
\label{fig11}
\vspace{-0.45cm}
\end{figure}

 \begin{figure}[!t]
\centering{\includegraphics[width=0.91\columnwidth] {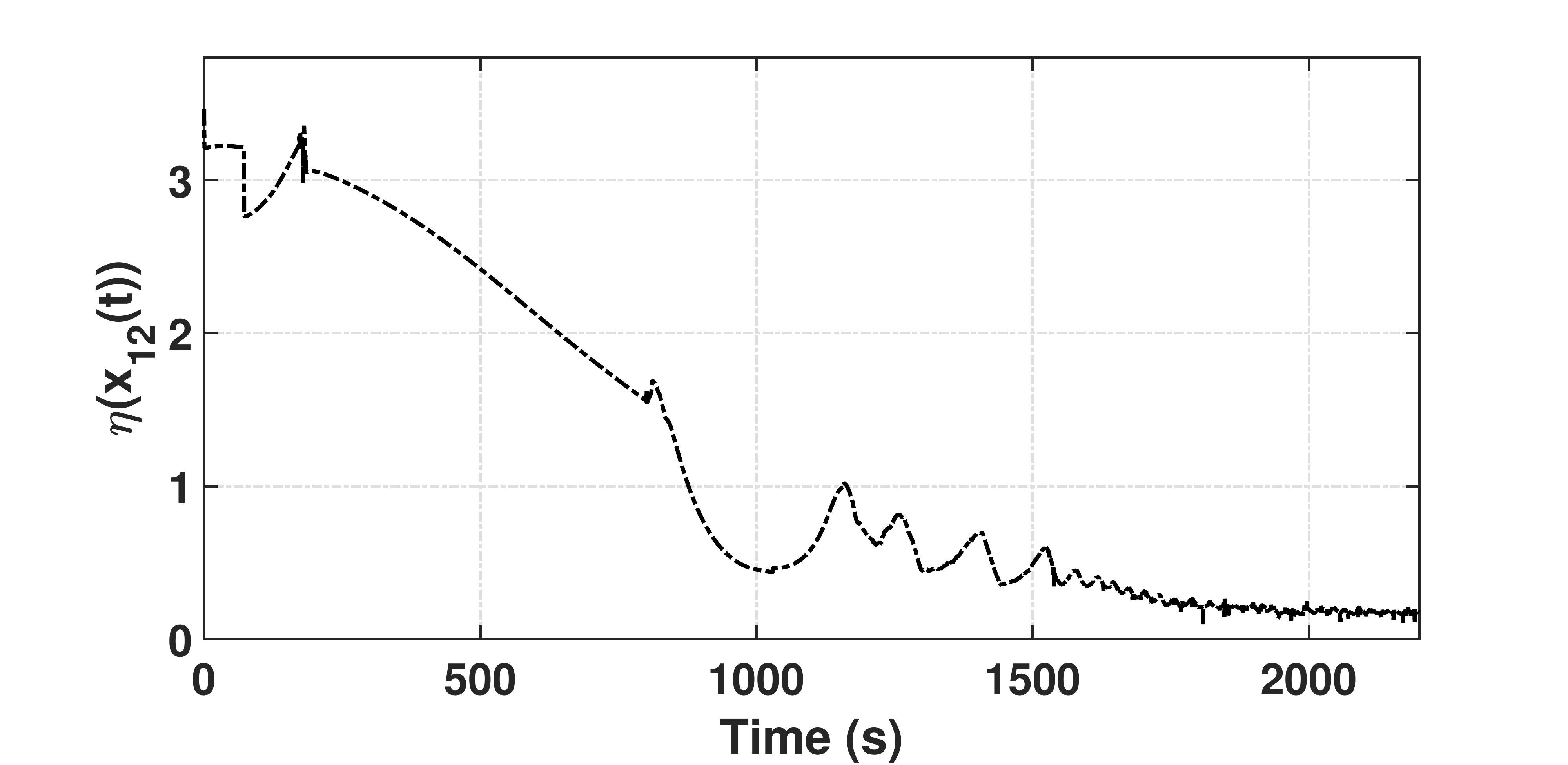}}
 \caption{Critical zone between agents 1 and 2 over time under the resilient QP in \eqref{QP2}.}
\label{fig12}
\vspace{-0.45cm}
\end{figure}


\begin{figure}[!t]
\centering{\includegraphics[width=0.92\columnwidth] {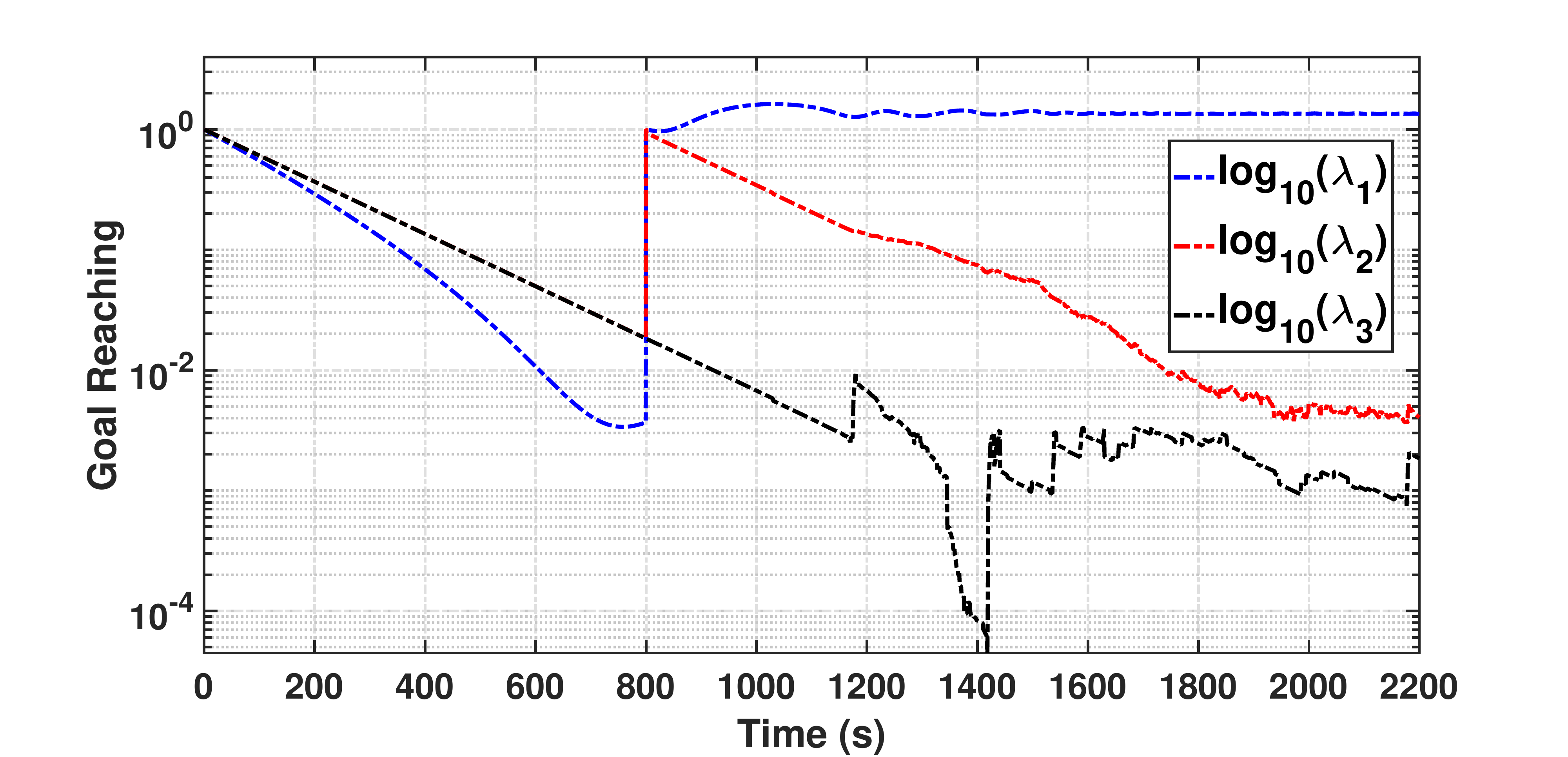}}
\caption{{Logarithmic Normalized resilient CLF for goal reaching behavior under adversarial chasing with resilient QP in \eqref{QP2} .}}
\label{fig10mod}
\vspace{-0.3cm}
\end{figure}



 \begin{figure}[!t]
\centering{\includegraphics[width=1\columnwidth] {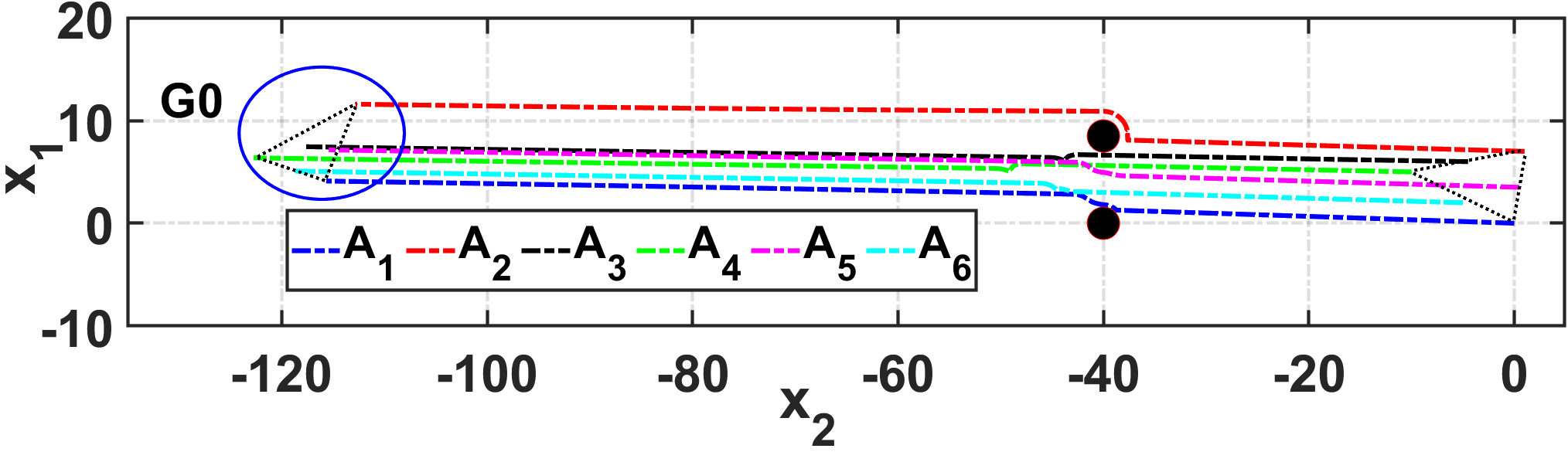}}
 \caption{Collaborative goal-reaching behavior of intact agents in formation in the absence of adversaries}
\label{fig14}
\vspace{-0.4cm}
\end{figure}

\begin{figure}[!t]
\begin{subfigure}{0.235\textwidth}
\centering{\includegraphics[width=0.85\columnwidth] {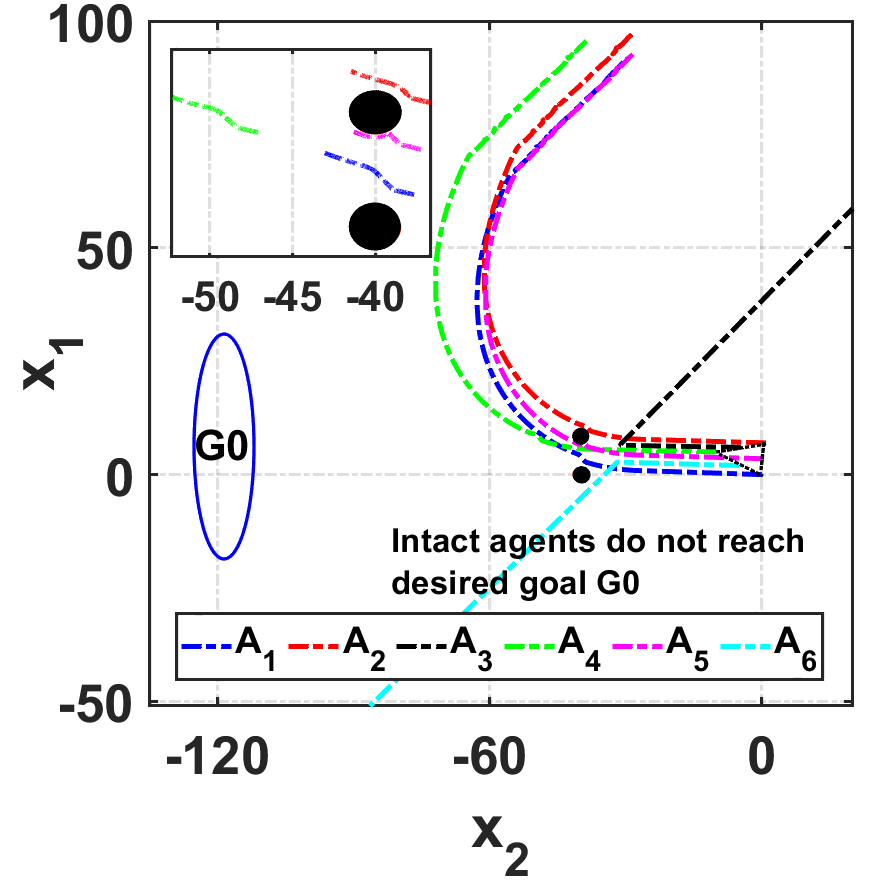}}
\caption{}
\label{fig13a}
\end{subfigure}
\begin{subfigure}{0.265\textwidth}
\centering{\includegraphics[width=0.95\columnwidth] {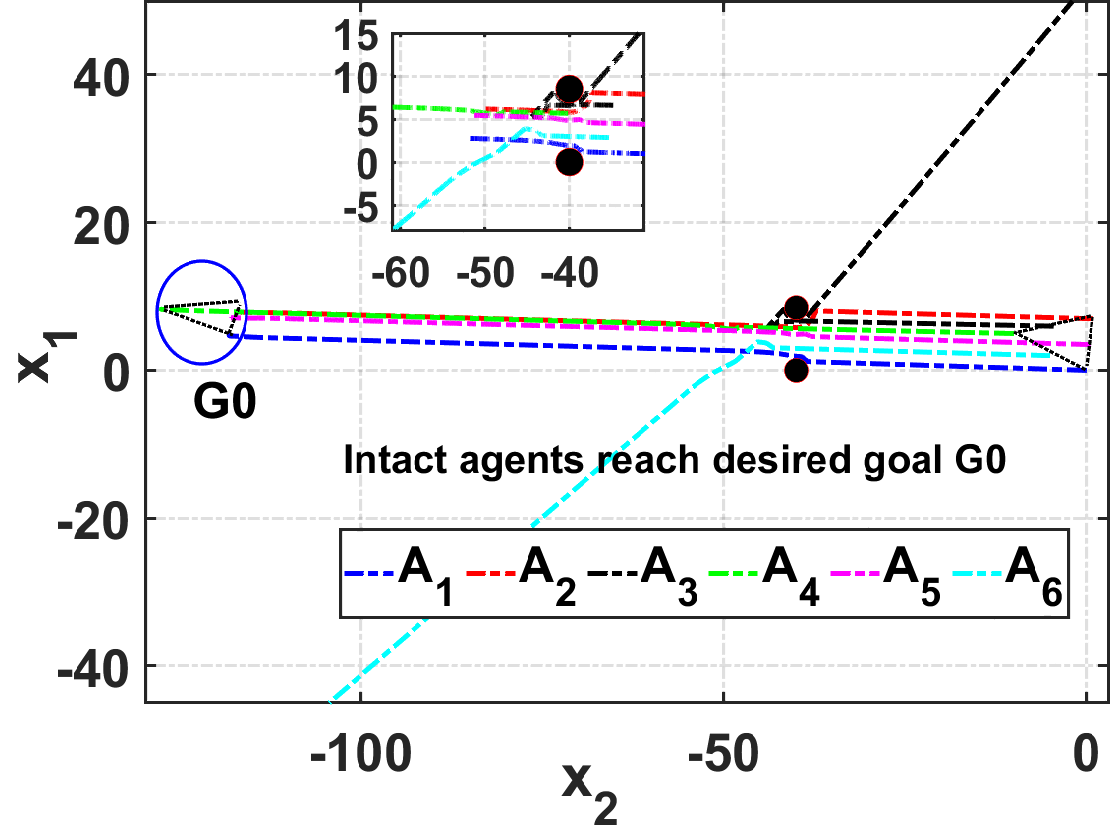}}
\caption{}
\label{fig13b}
\end{subfigure} 
\caption{(a) Collaborative goal-reaching behavior in the presence of adversarial agents 3 and 6. (b) Resilient collaborative goal-reaching behavior in the presence of adversarial agents 3 and 6. }
\label{fig13}
\end{figure}

Now, we consider Agent $1$ as adversarial and it performs adversarial chasing after $t>800$ with aim to achieve $\left\| x_{1}(t) -x_{2}(t) \right\| \to 0$ in some finite time.  Figure \ref{fig2b} shows the agent's behavior under adversarial chasing and violation of desired specifications $\phi$. We can see in Figure \ref{fig6} how the goal reaching behavior for Agent 1 starts growing after $t=800s$ due to adversarial chasing behavior. Similarly, Figure \ref{fig7} shows that the inter-agent safety violation happens at $t=1116s$ as the value of the safety behavior metric exceeds one, which implies that Agent 1 violates the safety constraint and hits the intact agent, i.e., Agent 2. Under adversarial chasing by Agent 1, one can see in Figure \ref{fig8} how the critical zone shrinks as the adversarial agent reaches close to intact Agent 2 and eventually goes to zero. Figure \ref{fig7} shows a proactive adversary detection happens at $t=907s$ by leveraging the concept of the critical zone for an inter-agent safety violation and it follows the result presented in Theorem 4. 

Now, based on the results of presented detection mechanism, we validate the efficacy of designed resilient QP in \eqref{QP2}. Figure \ref{fig10} illustrates the agents' behavior under adversarial chasing and under the resilient QP in \eqref{QP2}; one can see that Agent 1 keeps trying to hit Agent 2, but Agent 2 escapes from adversarial chasing based on the designed resilient inter-agent safety constraint in \eqref{QP2f}. Similarly, Figure \ref{fig11} shows that the resilient QP in \eqref{QP2} guarantees all-time inter-agent safety for intact agents even under adversarial chasing, as the safety metric is always less then one. Also, one can see that the critical zone in Figure \ref{fig12} does not shrink to zero as adversarial agent reaches close to but does not hit intact Agent 2. Figure \ref{fig10mod} illustrates the goal-reaching behavior for intact agent, i.e., Agents 2 and 3 eventually goes to zero and they reach their desired goal position $G_2$ and $G_3$. However, for Agent 1, the goal-reaching behavior grows after $t=800s$ due to adversarial chasing behavior, and Agent 1 never reaches the specified goal location $G_1$.







\noindent

\subsubsection{Case 2}
In the second case, we consider a multi-agent formation problem under some desired specifications, where the aim is to maintain formation among the set of collaborative agents and visit some regions, while maintaining safety constraints with collaborative goal reaching even in the presence of adversarial agents. In particular, we consider six agents with linearized unicycle dynamics and the desired goal location of formation centroid $G0=[-120 \,\,\,\, 7]^T$. Figure \ref{fig14}  shows the collaborative goal reaching for the agents without any adversarial agents, i.e., $\mathcal{A}_f=\emptyset$. One can see in Figure \ref{fig14} how agents maintain safety and reach desired goal location $G0$ over the desired time duration $t \in [0, 1000]$. Then, we consider the same scenario in the presence of multiple adversarial agents, i.e., Agents 3 and 6 act as adversarial agents for all $t > 400$ (both agents belong to class 2 type of adversarial agent as defined in Definition 2). It is shown in Figure \ref{fig13a} how adversarial agents mislead the collaborative goal reaching behavior and thus, intact agents do not reach the desired goal location $G0$, collectively over the time interval $t \in [0, 1000]$. Then, based on presented Algorithm 2 and Theorem 5, we detect the set of adversarial agents $\mathcal{A}_f$  and mitigate their effects in collaborative goal reaching. The Figure \ref{fig13b} illustrates that even in the presence of multiple adversarial agents, intact agents achieve the desired collaborative goal reaching behavior and reach the  goal location $G0$ over the desired time duration $t \in [0, 1000]$.

\section{Conclusions and Future Direction}
In this paper, we presented the proactive adversary detection mechanism and then designed a resilient control framework for multi-agent systems. In particular, first we analyzed agent's behaviors based on designed behavior metrics, and then designed proactive adversary detection mechanism based on the notion of the critical region for the system operation. The presented detection mechanism identified adversarial agents while ensuring all-time safety for normally behaving agents in the presence of adversarial agents. By leveraging the presented results for behavior analysis and adversary detection, we designed a resilient QP-based controller for multi-agent systems with desired safety and goal reaching constraints for intact agents, even in the presence of the adversarial agent. Finally, two case studies are presented to illustrate the efficacy of the presented theoretical contributions.

{A possible direction for future work is to extend the presented framework for resilience to more sophisticated models of adversarial actions instead of worst-case actions.}


\begin{IEEEbiography}[{\includegraphics[width=1in,height=1.25in,clip,keepaspectratio]{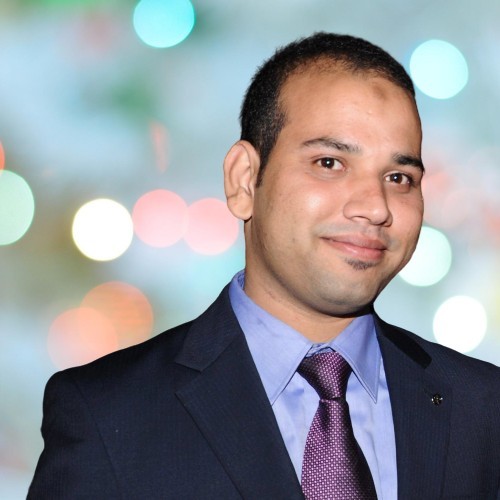}}]{{\bf Aquib Mustafa}}
 (S’17) received the Master’s degree from the Indian Institute of Technology Kanpur, Kanpur, India, in 2016. and the PhD degree from the Michigan State University, East Lansing, Michigan, USA, in 2020. He is currently working as a Postdoctoral Research Fellow with the Department of Aerospace Engineering, University of Michigan, Ann Arbor, MI, USA. His research interests include the Resilient control, Safety-critical systems, Reinforcement learning, and Multi-agent systems.
\end{IEEEbiography}

\begin{IEEEbiography}[{\includegraphics[width=1in,height=1.25in,clip,keepaspectratio]{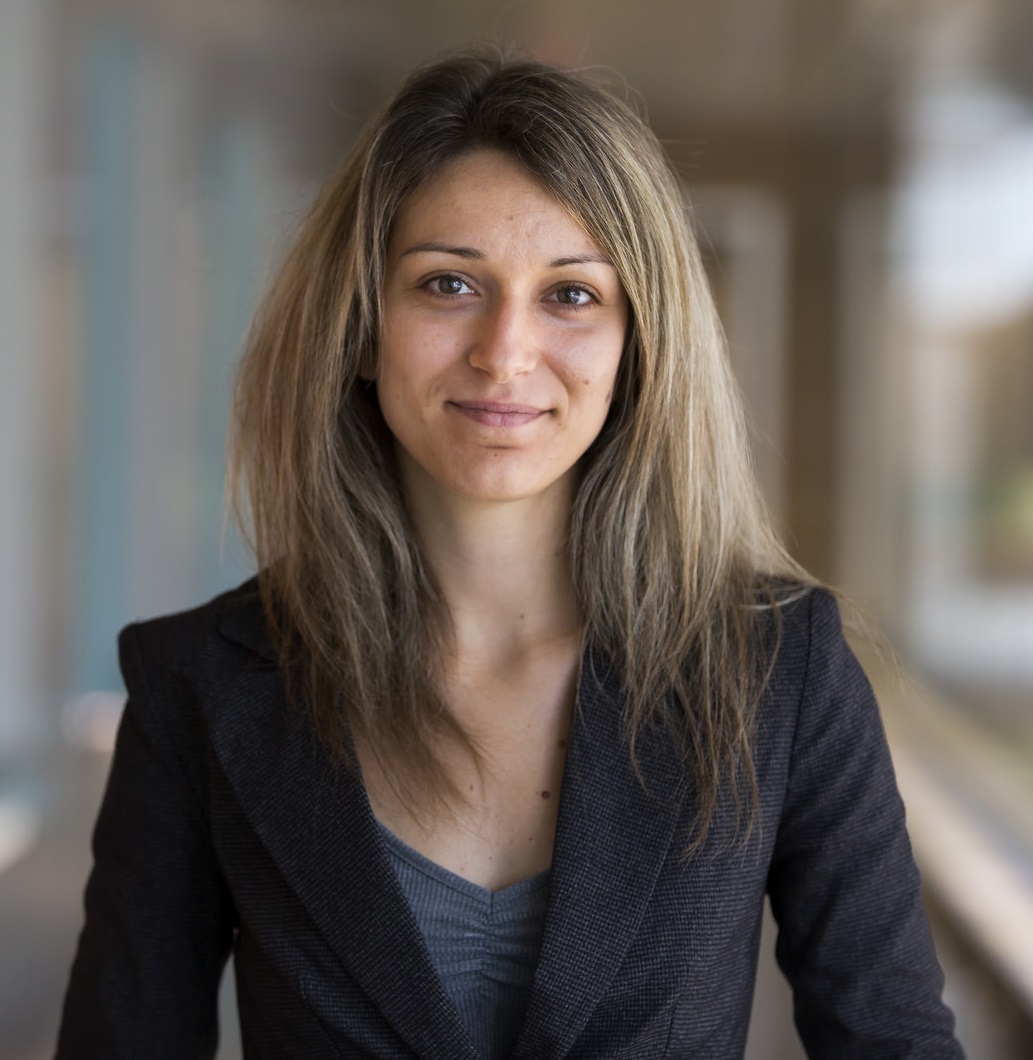}}]{{\bf Dimitra Panagou}} (Senior Member, IEEE) received
the Diploma and Ph.D. degrees in mechanical engineering from the National Technical University of Athens, Athens, Greece, in 2006 and 2012, respectively.\par
She is currently an Associate Professor with the Department of Aerospace Engineering, University of Michigan, Ann Arbor, MI, USA. Prior to joining the University of Michigan, she was a Postdoctoral Research Associate with the Coordinated Science Laboratory, University of Illinois at Urbana–Champaign, Champaign, IL, USA, a Visiting Research Scholar with the GRASP Lab, University of Pennsylvania, Philadelphia, PA, USA, and a Visiting Research Scholar with the Mechanical Engineering Department, University of Delaware, Newark, DE, USA. Her research interests include the fields of multiagent planning, control and estimation, with applications in safe and resilient robotic networks, autonomous multivehicle systems, and 
human–robot interaction.\par
Dr. Panagou was a recipient of the NASA 2016 Early Career Faculty Award, the AFOSR 2017 Young Investigator Award, and the NSF CAREER Award in 2020. She is a senior member of the AIAA.
\end{IEEEbiography}

\end{document}